\documentclass[a4paper,fleqn,usenatbib]{mnras}

\usepackage{newtxtext,newtxmath}

\usepackage[T1]{fontenc}
\usepackage{ae,aecompl}

\usepackage{graphicx}	
\usepackage{amsmath}	
\usepackage{amssymb}	
\usepackage{subfig}

\usepackage{array,booktabs,tabularx}
\newcolumntype{C}{>{\centering\arraybackslash}X}
\usepackage[usenames,dvipsnames,svgnames,table]{xcolor}

\newcommand{\lt}{<}

\title[Baryon Fraction of Haloes]{Using Artificial Neural Networks to Constrain the Halo Baryon Fraction during Reionization}

\author[D. Sullivan et al.]{David Sullivan$^{1}$\thanks{E-mail:
d.sullivan@sussex.ac.uk},  Ilian T. Iliev$^{1}$\thanks{E-mail:
I.T.Iliev@sussex.ac.uk} and Keri L. Dixon$^{1}$\\
$^{1}$Astronomy Centre, Department of Physics and Astronomy, University of Sussex, Brighton, BN1 9QH, U.K.}

\date{Accepted XXX. Received YYY; in original form ZZZ}

\pubyear{2017}

\begin{document}
\label{firstpage}
\pagerange{\pageref{firstpage}--\pageref{lastpage}}
\maketitle

\begin{abstract}
Radiative feedback from stars and galaxies has been proposed as a potential solution to many of the tensions with simplistic galaxy formation models based on $\Lambda$CDM, such as the faint end of the UV luminosity function. The total energy budget of radiation could exceed that of galactic winds and supernovae combined, which has driven the development of sophisticated algorithms that evolve both the radiation field and the hydrodynamical response of gas simultaneously, in a cosmological context. We probe self-feedback on galactic scales using the adaptive mesh refinement, radiative transfer, hydrodynamics, and $N$-body code {\sc ramses-rt}. Unlike previous studies which assume a homogeneous UV background, we self-consistently evolve both the radiation field and gas to constrain the halo baryon fraction during cosmic reionization. We demonstrate that the characteristic halo mass with mean baryon fraction half the cosmic mean, $M_{\mathrm{c}}(z)$, shows very little variation as a function of mass-weighted ionization fraction. Furthermore, we find that the inclusion of metal cooling and the ability to resolve scales small enough for self-shielding to become efficient leads to a significant drop in $M_{\mathrm{c}}$ when compared to recent studies. Finally, we develop an Artificial Neural Network that is capable of predicting the baryon fraction of haloes based on recent tidal interactions, gas temperature, and mass-weighted ionization fraction. Such a model can be applied to any reionization history, and trivially incorporated into semi-analytical models of galaxy formation.
\end{abstract}




\section{Introduction}
\label{sec:intro}

During the first billion years after the Big Bang, the large-scale cosmic web of structures we see today began to form, followed by the first stars and galaxies, which brought an end to the Dark Ages \citep{Rees:1999aa}. These first luminous sources are thought to be prime candidates to drive cosmic reionization, a major phase transition of the Universe, from a neutral intergalactic medium (IGM) following recombination, to the ionized state it still remains in today. This period is known as the Epoch of Reionization (EoR), and the wide range of physical processes which drive it encapsulate several areas of research, from cosmology and galaxy formation to radiative transfer (RT) and atomic physics. Even with the wealth of present-day observational information at our disposal, these processes are not fully understood, and still less so in the high-redshift Universe.

Photoheating of the IGM by stars (and likely also by quasars and other radiation sources) raises the fully-ionized gas temperature to $T_{\mathrm{IGM}} \sim 10^{4}\ \mathrm{K}$, preventing the cooling and adiabatic collapse of gas into low-mass dark matter (DM) haloes. This quenching was originally discussed by \cite{Doroshkevich:1967aa} and investigated by \cite{Couchman:1986aa} within the cold dark matter (CDM) framework, and may explain the discrepancy between the observed faint end galaxy luminosity function and the prediction from CDM, known as the missing satellites problem \citep{Moore:1999aa}. Furthermore, the smallest dwarf galaxies can be photoevaporated entirely: three dimensional RT simulations that probe the propagation of ionization fronts (I-fronts) show that as supersonic R-type fonts encounter dense neutral gas in (mini) haloes, they quickly decelerate to D-type fronts, preceded by shock waves which blow the gas back into the IGM as an ionized supersonic wind \citep{Shapiro:2004aa, Iliev:2005ab}. This occurs on time-scales of hundreds of millions of years, reducing the baryon fraction to a few percent (or less) of the total halo mass. As a consequence, the resulting increased photon consumption rate slows the propagation of global ionization fronts and extends reionization \citep{Iliev:2005aa}.

Recent studies have attempted to quantify the reduction of baryons in low-mass haloes using semi-analytical-models (SAMs) \citep[e.g][]{Babul:1992aa, Efstathiou:1992aa, Shapiro:1994aa, Nagashima:1999aa, Benson:2002aa, Benson:2002ab, Somerville:2002aa} and full 3D cosmological simulations \citep[e.g][]{Quinn:1996aa, Weinberg:1997aa, Gnedin:2000ab, Hoeft:2006aa, Okamoto:2008aa, Noh:2014aa}. In the absence of a photoionizing background, the baryon fraction $f_{\mathrm{b}} = M_{\mathrm{b}}/M_{\mathrm{tot}}$ of haloes scatters around the cosmic mean $\langle f_{\mathrm{b}} \rangle = \Omega_{\mathrm{b}}/\Omega_{\mathrm{M}}$. Supernovae (SNe) are not expected to significantly effect $f_{\mathrm{b}}$, as even the most efficient scenario (Sedov blast wave) can only accelerate gas to a few hundred km$\rm s^{-1}$, while the escape velocity of a Milky Way sized halo is $v_{\mathrm{esc}} \sim 700\ \mathrm{km/s}$ (and higher at high-redshift, where haloes are denser and more compact). As a result, SNe feedback is largely important \emph{only} for dwarf galaxies \citep{Dubois:2008aa, Roskar:2014aa}, however, such events become increasingly rare towards lower masses. For a single stellar population, the available radiative energy of massive stars outweighs the cumulative feedback budget from galactic winds and SNe Type II explosions by a factor of $\sim 100$ \citep{Roskar:2014aa}. While this depends on the interstellar medium (ISM) being able to efficiently absorb radiation, it suggests that radiative feedback could be a crucial ingredient in the regulation of star formation in galaxies.

Currently, SAMs rely on hydrodynamical simulations to constrain $f_{\mathrm{b}}$ in their galaxy formation models. Typically, one matches the halo mass with a baryon fraction half the cosmic mean, known as the \emph{Characteristic mass}, $M_{\mathrm{c}}$. This scale sets the mid-point in the transition from baryon poor to baryon rich, thus can be used to generate simple fitting functions, which tightly constrain $f_{\mathrm{b}}$ as a function of halo mass. Prior to this work, $M_{\mathrm{c}}$ was thought to be set by the Jeans mass (or the \emph{filtering scale}, which takes into account the halo mass assembly; \citealt{Shapiro:1994aa, Gnedin:1997aa, Gnedin:2000aa}), evaluated at the mean density of the Universe. The filtering scale was found to overshoot $M_{\mathrm{c}}$ during the EoR by \cite{Hoeft:2006aa} and \cite{Okamoto:2008aa}, who instead argued that gas can accrete only if the haloes equilibrium temperature (where photoheating balances cooling), evaluated at over-densities in the range $60 \leq \Delta \leq 1000$, is greater than the virial temperature. However, even these studies were found to overshoot $M_{\mathrm{c}}$, as the true bottleneck for collapse occurs at densities that are an order of magnitude lower than the virial density and an order of magnitude above the cosmic mean density. At these densities, gas is not yet able to cool efficiently, while once it reaches the virial density, it will almost certainly continue to collapse as it can radiate away its energy efficiently \citep{Noh:2014aa}. Clearly, there is a great deal of uncertainty surrounding the physical processes which set $M_{\mathrm{c}}$.

While computational algorithms have significantly improved since these studies, little advancement has been made on constraining the halo baryon fraction. Therefore, we are motived to revisit this in the context of fully self-consistent radiation hydrodynamical (RHD) simulations. We compare our results to the predictions of \cite{Okamoto:2008aa} and \cite{Hoeft:2006aa}, developing and training an artificial neural network (ANN), which reproduces our simulations well. This model is independent of reionization history and can be incorporated into SAMs to self-consistently compute $f_{\mathrm{b}}$, better capturing the scatter.

This paper is arranged as follows. The details of our methods, star formation calibration, and SNe feedback scheme are outlined in Section~\ref{sec:method}. Our simulation set-up, along with key parameters and choice of halo finder is discussed in Section~\ref{sec:cosmo_sims}, where we also study radiative feedback on galaxies using a statistical analysis of haloes. In this section we also introduce our results for the halo baryon fraction for three different reionization histories. We describe our ANN architecture in Section~\ref{sec:ANN}, along with our choice of back-propagation algorithm and adaptive learning rate.  We conclude this section by presenting our constraints on $M_{\mathrm{c}}(z)$, and use our ANN to predict this quantity based on our simulation results and a test model where we remove strong tidal forces. Finally, we discuss our results in the context of the literature in Section~\ref{sec:discussion} and present our final conclusions in Section~\ref{sec:conclusions}.

\section{Method}
\label{sec:method}
\subsection{Radiation Hydrodynamics Code}

To investigate the impact of feedback during the EoR in a cosmological context, we use the Eulerian adaptive mesh refinement (AMR) code {\sc ramses} (\cite{Teyssier:2002aa}; ver. 3). Basic grid elements are known as \emph{octs}, which each belong to a particular level in the AMR hierarchy, $\ell$, and are comprised of $2^{\mathrm{ndim}}$ cells, where ndim denotes the number of spatial dimensions (3 here). The code is based on the "Fully Threaded Tree" structure of \cite{Khokhlov:1998aa}, where refinement is performed on a cell-by-cell basis following the user desired criteria. The hydrodynamics solver employs a second-order Godunov scheme, proven to capture shocks and accurately follow the thermal history of the fluid with hydrodynamical states reconstructed on cell interfaces using a MinMod method and advanced using the Harten-Lax-van-Leer contact wave Riemann solver (HLLC; \citealt{Toro:1994aa}). The Poisson equation is solved using an adaptive particle-mesh method, with the collisionless $N$-body system described by the Vlasov-Poisson equations. Gas is allowed to cool through atomic excitation/de-excitation down to $10^{4}$ K and further to $\sim 100$ K via metal fine-line transitions

We use the multi-group RT module {\sc ramses-rt}, developed by \cite{Rosdahl:2013aa}, to compute the photoionization of three species: \ion{H}{i}, \ion{He}{i}, and \ion{He}{ii}. The code uses a moment-based scheme to solve the RT equation for three photon groups (one for each species) using a first-order Godunov method with the M1 closure for the Eddington tensor \citep{Aubert:2008aa, Aubert:2010aa}. By treating the photons as a fluid in this way, the code makes natural use of the existing solvers implemented in {\sc ramses} to fully couple the radiation with the gas via photoionization and photoheating and a set of non-equilibrium chemistry equations for the three species above. This scheme has the advantage over ray-tracing methods that the RT step is independent of the number of sources, ideal for cosmological simulations of galaxy formation. The production rate of ionizing photons varies with time for a given IMF, with the majority of photons released within 5 Myr from the birth of a star particle. We adopt the \emph{on-the-spot} approximation, where UV photons emitted from recombinations are assumed reionize an atom in the same grid cell, hence case B recombination rates are used when computing the gas cooling rate.

\subsection{Star Formation}

The highest resolution simulations of galaxy formation in cosmological contexts are still limited to parsec scales, far larger than the scale at which the formation of a single star takes place. Therefore, a sub-grid recipe is implemented in {\sc ramses}, which follows a standard Schmidt law in each cell:
\begin{equation}
\dot{\rho}_{*} = \epsilon_{\mathrm{ff}}\rho/t_{\mathrm{ff}},
\end{equation}
where $\epsilon_{\mathrm{ff}} = 0.01$ is the star formation efficiency per free fall time $t_{\mathrm{ff}} = [3\pi/(32G\rho)]^{1/2}$, and $G$ is the gravitational constant. Cells are considered star-forming when they meet the predefined density and temperature criteria:
\begin{align}
\begin{split}
\label{eqn:sf_thresh}
&n_{*} \geq 8\ \textrm{cm}^{-3}\\
&T/\mu \leq 2\times10^{4}\ \textrm{K},
\end{split}
\end{align}
where $\mu$ is the mean molecular weight of the gas. When a cell becomes eligible, $N_{*}$ collisionless star particles are formed stochastically, with probability drawn from a Poisson distribution:
\begin{equation}
P(N_{*}) = \frac{\lambda_{\mathrm{P}}}{N_{*}!} \exp (-\lambda_{\mathrm{P}}),
\end{equation}
with mean value
\begin{equation}
\lambda_{\mathrm{P}} = \left( \frac{\rho \Delta x^{3}}{m_{*}} \right) \frac{\Delta t}{t_{\mathrm{ff}}},
\end{equation}
where $\Delta x$ is the minimum cell width and $\Delta t$ is the simulation time step \citep{Rasera:2006aa}. The stellar particle mass is then determined as an integer multiple of $m_{*} = n_{*} \Delta x^{3} / (1 + \eta_{\mathrm{SNe}} + \eta_{\mathrm{w}})$, where the parameters $\eta_{\mathrm{ \mbox{{\tiny SNe}} }}$ and $\eta_{\mathrm{w}}$ are discussed in Section~\ref{sec:sne_feedback}.

\subsection{Calibration}
\label{sec:calibration}
\begin{figure}
	\includegraphics[width=\columnwidth]{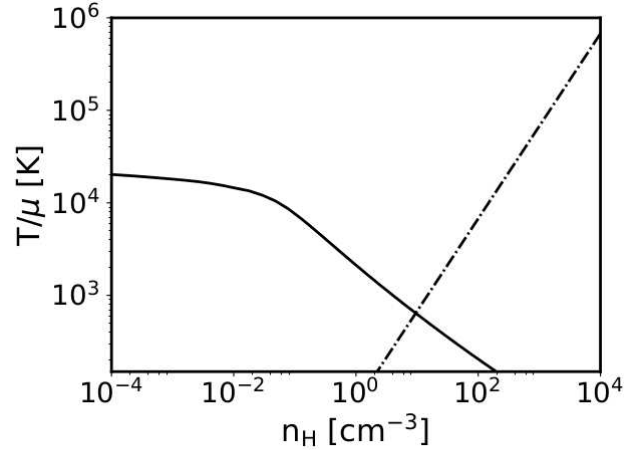}
    \caption{ Temperature-density relation (solid) and intersection with our polytropic temperature floor (dot-dashed) used to calibrate star formation criteria. The solid line shows the trajectory in the temperature-density plane for a single cell at solar metallicity, as calculated using the {\sc ramses} cooling module.}
\label{fig:jeans_calib}
\end{figure}

An artificial `Jeans pressure' is imposed on the gas to prevent numerical fragmentation below the Jeans scale of fine grid cells \citep{Truelove:1997aa}, which can be written as:
\begin{equation}
\lambda_{\mathrm{J}} = \sqrt{ \frac{\pi c_{\mathrm{s}}}{G\rho} } = 16\ \textrm{pc} \left(\frac{T}{1\ \textrm{K}} \right)^{ \frac{1}{2} } \left(\frac{n_{\mathrm{H}}}{1\ \textrm{cm}^{-3}} \right)^{ - \frac{1}{2} },
\label{eqn:jeans_length}
\end{equation}
where $c_{\mathrm{s}} = \sqrt{\gamma k_{\textrm{B}} T / m_{\textrm{p}}}$ is the sound speed, and we assume a ratio of specific heats $\gamma = 5/3$ for a monoatomic gas. Imposing the condition that the Jeans length be resolved on $N$ cells gives rise to a temperature floor, which must satisfy \citep{Rosdahl:2015aa}:
\begin{equation}
\frac{T}{1 \textrm{K}} \geq \frac{n_{\mathrm{H}}}{1\ \textrm{cm}^{-3}} \left(\frac{N \Delta x}{16\ \textrm{pc}} \right)^{2}.
\end{equation}
We apply this in the form of an effective temperature function (added to the cell temperature) with polytropic equation of state:
\begin{equation}
T_{\mathrm{J}} = T_{0} \left( \frac{n_{\mathrm{H}}}{n_{*}} \right)^{g_{*} - 1},
\end{equation}
where a value $T_{0} = 540\ \textrm{K}$ ensures star-forming gas is always resolved by at least six cells in our simulations and $g_{*}$ is the polytropic index (set to 2 in our models). Note that as this pressure floor is non-thermal, the gas temperature $T$ can fall below this floor, therefore, the quantity $T - T_{\mathrm{J}}$ must satisfy the temperature criterion in equation~(\ref{eqn:sf_thresh}). In Fig.~\ref{fig:jeans_calib}, we show our polytropic equation of state and its intersection with the equilibrium cooling curve, which we calculate using the cooling module from {\sc ramses}. The cooling module implements thermal Bremsstrahlung, ionization, recombination, dielectric recombination, and metal-line cooling, as-well as radiative and Compton heating. We compute the temperature-density relation for a single cell at solar metallicity, and proceed to calibrate by adjusting $n_{*}$ and $T_{0}$ such that the intersect point remains fixed for a given resolution.

\subsection{Supernovae Feedback}
\label{sec:sne_feedback}

We incorporate SNe using the kinetic feedback implementation of \cite{Dubois:2008aa}, which models the Type II SNe explosion by increasing the kinetic energy of the surrounding gas within a spherical blast wave of radius $r_{\mbox{{\tiny SNe}}}$. Once a cell becomes star-forming, gas mass is depleted and converted into star particles such that
\begin{equation}
(\Delta m_{\mathrm{g}})_{\mathrm{SF}} = m_{*} (1 + \eta_{ \mbox{{\tiny SNe}} } + \eta_{\mathrm{w}}),
\end{equation}
where $m_{*}$ is the star particle mass, $\eta_{ \mbox{{\tiny SNe}} }$ is the SNe mass fraction, and $\eta_{\mathrm{w}}$ is the mass-loading factor (i.e. the amount of gas entrained within the SNe ejecta; \citealt{Springel:2003aa}). A value of $\eta_{ \mbox{{\tiny SNe}} } = 0.2$ is chosen, which roughly corresponds to a \cite{Chabrier:2003aa} stellar IMF. Observational evidence suggests the mass-loading factor varies with stellar mass \citep{Martin:2005aa, Rupke:2005aa, Weiner:2009aa, Martin:2012aa, Newman:2012aa, Chisholm:2015aa, Arribas:2014aa}; however, a value $\eta_{w,\mathrm{max}} = 10$ is acceptable and therefore used here. The SNe explosion is treated as a single event 10 Myr after the formation of the star particle, releasing kinetic energy:
\begin{equation}
E_{\mathrm{d}} = \eta_{ \mbox{{\tiny SNe}} } \frac{m_{*}}{M_{ \mbox{{\tiny SNe}} }}E_{ \mbox{{\tiny SNe}} }
\end{equation}
into the neighbouring gas cells, where $M_{ \mbox{{\tiny SNe}} }$ and $E_{ \mbox{{\tiny SNe}} }$ are fixed for a typical Type II SNe explosion (10 M$_{\odot}$ and $10^{51}$ erg, respectively). In practice, the final mass-loading factor can be as low as zero, depending on the density of gas in neighbouring cells at the point of explosion. The most important parameter in the model is the radius over which the fluid variables are modified, $r_{ \mbox{{\tiny SNe}} }$, which we choose to be 200 pc and is consistent with the predicted size of galactic super-bubbles \citep{McKee:1977aa}. While this scheme is not as sensitive to the over-cooling problem as thermal dump methods, it is also not immune: kinetic energy is converted into heat via shocks immediately after the SNe explosion. Given the typically high Mach number, ambient density, and limited numerical resolution, this heat can be efficiently radiated away, reducing the overall impact of SNe feedback \citep{Kimm:2014aa}. As SNe are not thought to play an important role in the stripping of baryons from haloes, we do not expect this to significantly effect our results \citep{Dubois:2008aa, Roskar:2014aa}.

\section{Cosmological Simulations}
\label{sec:cosmo_sims}

We assume a geometrically flat $\Lambda$CDM Universe consistent with a WMAP-7 cosmology \citep{Komatsu:2011aa} with cosmological parameters: $\Omega_{\mathrm{M}} = 0.272$, $\Omega_{\Lambda} = \Lambda_{0}/(3 H_{0}^{2}) = 0.728$, $\Omega_{\mathrm{b}} = 0.045$, $h \equiv H_{0}/(100$ km s$^{-1}$ Mpc$^{-1}) = 0.702$, and $\sigma_{8} = 0.82$, where the symbols take their usual meaning. Initial conditions are generated using the {\sc music} code \citep{Hahn:2013aa} at redshift $z = 100$ in a comoving volume of $\sim 5.7$ Mpc on-a-side. Variations in these cosmological parameters are not expected to largely influence our results.

The coarse grid level is set to $\ell_{\mathrm{min}} = 8$ in the AMR hierarchy for all simulations, resulting in $(2^{\ell_{\mathrm{min}}})^{3} = (256)^{3}$ DM particles and coarse grid cells, such that our DM particle mass $M_{\mathrm{DM}} \sim 3.4 \times 10^{5}\ \mathrm{M}_{\odot}$. We set the maximum level of refinement $\ell_{\mathrm{max}} = 18$, such that our effective spatial resolution is $\Delta x_{\mathrm{min}} = 21.72$ pc comoving and our stellar particle mass is $m_{*,\mathrm{min}} \sim 2 \times 10^{3} M_{\odot}$. We adopt uni-grid initial conditions for the gas and dark matter (i.e.., no zoom regions).

Dark matter (sub) haloes are catalogued using the {\sc rockstar} phase-space halo finder \citep{Behroozi:2013ab}, while merger trees are constructed using the {\sc consistent-trees} algorithm \citep{Behroozi:2013ac}. Particles are first divided into 3D Friends-of-Friends (FOF) groups, which are analysed in 6D phase-space to give robust, grid- and shape-independent haloes. Haloes that are not gravitationally bound are removed and properties are computed using the viral overdensity definition of \cite{Bryan:1998aa}.


Photoionization of both hydrogen and helium is followed via the coupling of UV photons with the gas, assuming a \cite{Bruzual:2003aa} stellar population for an instantaneous Chabrier IMF \citep{Chabrier:2003aa}. We perform a suite of RHD simulations with stellar particle escape fractions, $f^{*}_{\mathrm{esc}} = 1.5,\ 2, and 5$, which is multiplied by the total number of photons integrated from our spectral energy distribution (SED) when emitting from star particles as listed in Table~\ref{tab:sim_names}.  Due to computational constraints, we adopt a reduced speed of light, $f_{\mathrm{c}} = 1/100$, which relaxes the Courant-condition for the radiation fluid and, thus, reduces the number of radiation sub-cycles necessary to reach thermal convergence \citep{Gnedin:2001ab}. We choose simulation RT2 as our fiducial model, as its reionization history falls between our RT1.5 and RT5 models. 

\begin{table}
\centering
\begin{tabularx}{\linewidth}{*{3}{C}}
    \toprule
    Name & Radiative Transfer & $f^{*}_{\mathrm{esc}}$\\\midrule
    RT1.5 & Yes & 1.5\\
    RT2 (Fiducial) & Yes & 2.0\\
    RT5 & Yes & 5.0\\
    HD & No & N/A
    \\\bottomrule
\end{tabularx}
 \caption{ List of simulations. We denote all RT simulations as RTX, where X refers to the choice of stellar escape fraction, $f^{*}_{\mathrm{esc}}$. For reference, we include a model with RT disabled (i.e.., SNe feedback only), which we denote HD. We choose RT2 as our fiducial model, as its reionization history falls between our other two models (see Fig.~\ref{fig:reion_hist}). }
\label{tab:sim_names}
\end{table}

Several authors have investigated the evolution and geometry of the EoR on large-scales, which accurately capture the statistics of this process \citep[e.g][]{Ciardi:2003aa, Iliev:2006aa, Mellema:2006aa, Zahn:2007aa, McQuinn:2007aa, Iliev:2007ab, Mesinger:2007aa, Geil:2008aa, Choudhury:2009aa, Dixon:2016aa}, and the general consensus is that reionization progresses inside-out, with high-density regions being ionized earlier, on average, than the voids. In such a scenario, the complex interplay between the growth of \ion{H}{ii} regions and the formation of structure in the Universe is crucial; while I-fronts from the first sources can quickly escape into voids, the exponential growth of structure during this extended process ensures that there are always new stars forming in the high-density peaks, which overwhelm reionization. The mass-weighted neutral fraction therefore remains steadily lower than the volume-weighted quantity throughout. To properly follow the inside-out progression of reionization, large-volumes are required which fully capture the statistics of structure formation \citep{Mellema:2006ab, Iliev:2006aa, Iliev:2014aa}.

\begin{figure}
	\includegraphics[width=\columnwidth]{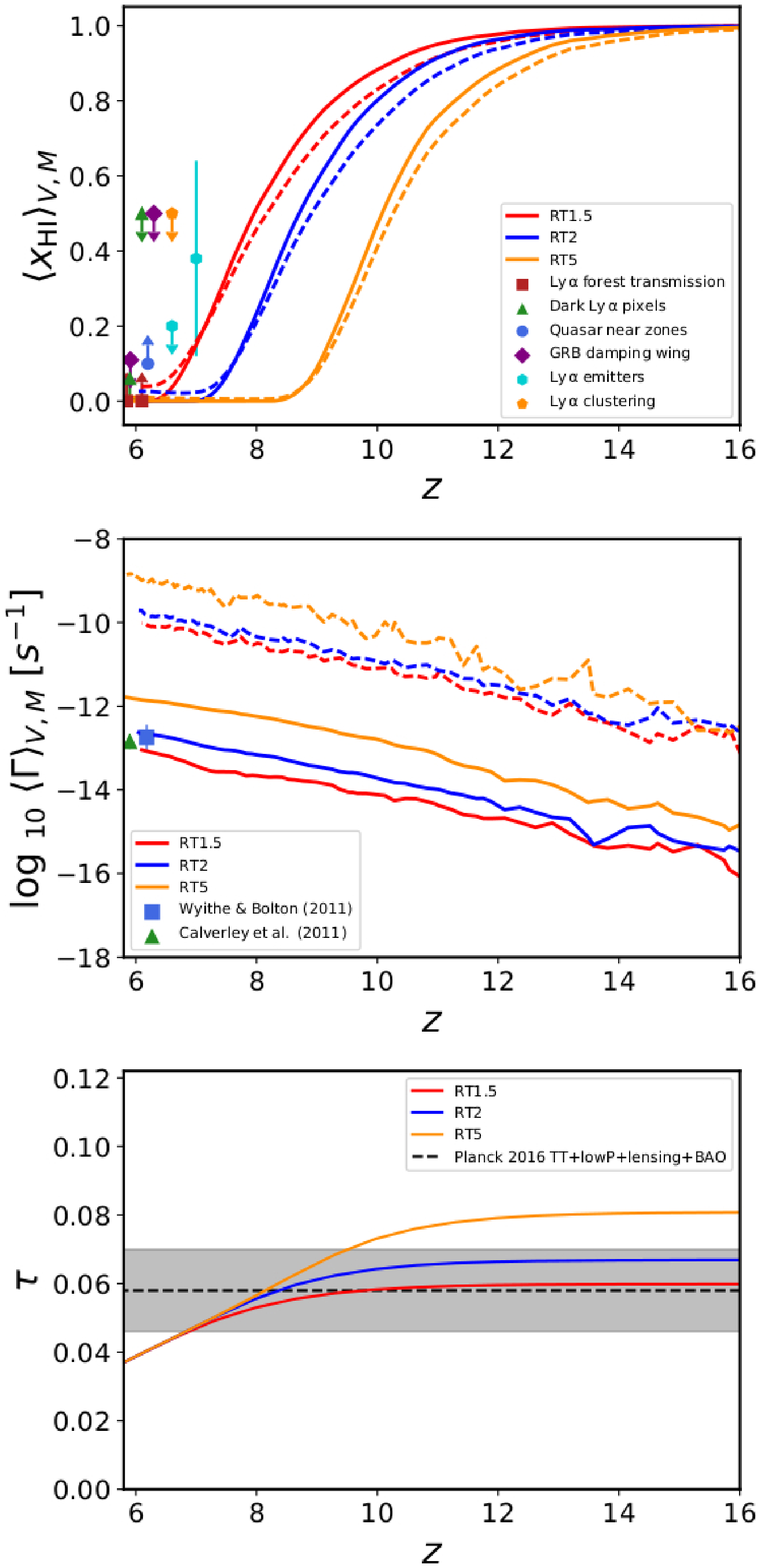}
\caption{ \emph{Top panel:} the volume/mass-weighted (solid/dahsed respectively) mean neutral fraction of hydrogen for all RTX simulations listed in Table~\ref{tab:sim_names}, compared to observational constraints from Ly $\alpha$ forest transmission (squares, red; \protect\citealt{Fan:2006aa}), dark Ly $\alpha$ forest pixels (triangles, green; \protect\citealt{McGreer:2011aa, McGreer:2015aa}), quasar near zones (circles, blue; \protect\citealt{Schroeder:2013aa}), GRB damping wind absorption (diamonds, voilet; \protect\citealt{McQuinn:2008aa, Chornock:2013aa}), and decline in Ly $\alpha$ emitters (hexagons, cyan; \protect\citealt{Ota:2008aa, Ouchi:2010aa}) following from \protect\cite{Robertson:2015aa}. \emph{Middle panel:} The mean volume/mass-weighted photoionization rate, compared to the volume-weighted observations of \protect\cite{Calverley:2011aa} and \protect\cite{Wyithe:2011aa} as the (green) triangle and (blue) square, respectively. \emph{Bottom panel:} The integrated electron-scattering optical depth compared to the PlanckTT+lowP+lensing+BAO 2016 results (thin black horizontal line) and the 1$\sigma$ error interval \citep[shaded region;][]{Planck-Collaboration:2016ab}. }
\label{fig:reion_hist}
\end{figure}

Early small-volume simulations predict an outside-in progression (e.g \citealt{Gnedin:2000ab}), with I-fronts quickly escaping into voids and slowly eating their way through filaments. Due to the limited volume, these models do not form as many new sources capable of overwhelming reionization; therefore, the opposite effect is found, whereby the volume-weighted neutral fraction lags behind the mass-weighted value. In Fig.~\ref{fig:reion_hist}, we show the evolution of both the volume-weighted and mass-weighted (solid and dashed, respectively) neutral fraction (top panel), photoionization rate ($\Gamma$; middle panel), and the integrated electron-scattering optical depth ($\tau$; bottom panel) for each of our RT models, along with relevant constraints on each quantity.

\begin{figure*}
	\includegraphics[width=\textwidth]{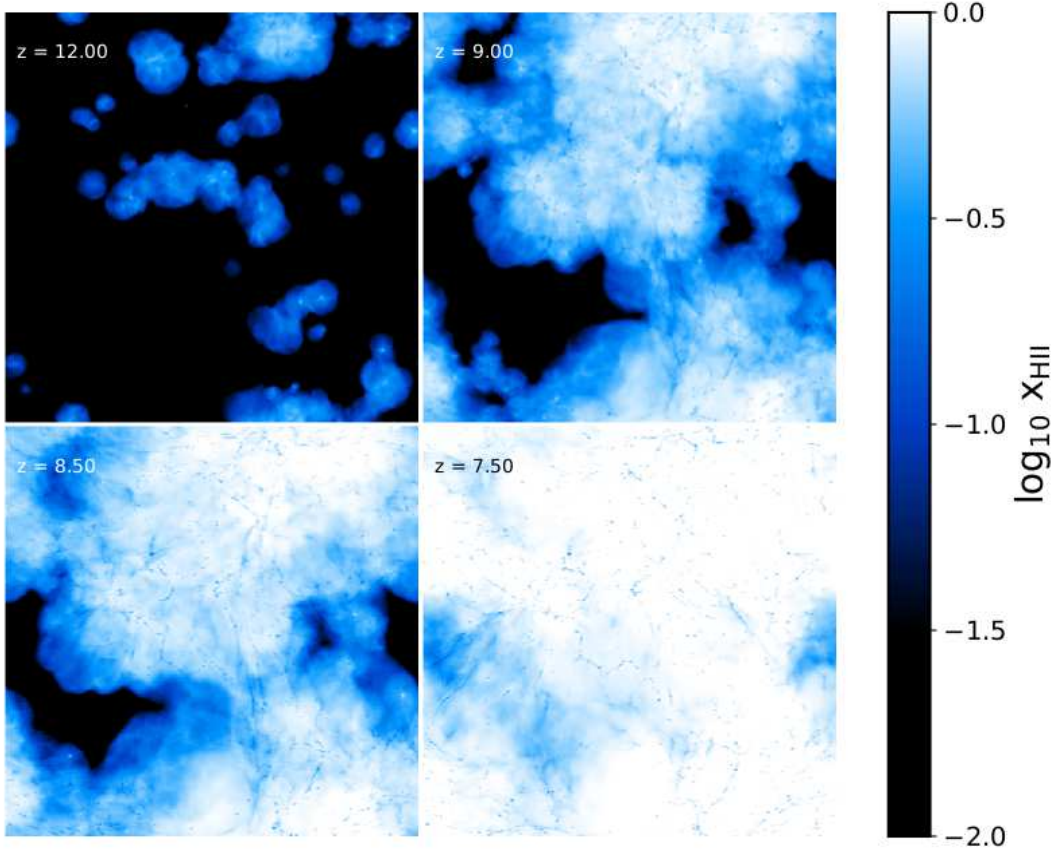}
    \caption{ Mass-weighted projections of the hydrogen ionization fraction, \ion{H}{ii}, at redshifts 12, 9, 8.5, and 7.5 in our fiducial RT2 model. Each panel shows the full projection of the (5.7 Mpc)$^{3}$ volume, projected along the z-axis. At $z = 12$, we see the formation of \ion{H}{ii} regions around the first stars in the universe, which grow and eventually overlap completing the EoR. Our model exhibits an inside-out progression, consistent with the literature and discussed in the text, whereby voids are the last regions to undergo reionization (as seen in the final panel; $z = 7.5$). }
    \label{fig:xHII_mw_proj}
\end{figure*}

For all models the mass-weighted neutral fraction is consistently below the volume-weighted quantity during the EoR, which is characteristic of an inside-out progression. This is depicted in Fig.~\ref{fig:xHII_mw_proj}, where we show mass-weighted projections of the fraction of ionized hydrogen, $\mathrm{x}_{\ion{H}{ii}}$, of the full volume at redshifts $z = 12$, $9$, $8.5$ and $7.5$. The initial growth of \ion{H}{ii} regions stems from the high density peaks (top left panel), and gradually grow before percolating and the completion of hydrogen reionization. This occurs shortly after $z = 7.5$, where in the bottom right panel we can still see a few pockets of neutral gas in underdense regions. Ionization fronts sweep through these regions last, leading to a steep drop in the volume-weighted neutral fraction at the end of the EoR as they make up a large fraction of the volume (in contrast to the mass-weighted value, as the solid lines converge/overtake the dashed lines).

Our fiducial model shows good agreement with constraints on $\Gamma$ and $\tau$, however is in slight tension with neutral fraction constraints from Lyman-$\alpha$ emitters. While this tension is not entirely relieved by our RT1.5 model, it shows much better agreement for both the neutral fraction and PlanckTT+lowP+lensing+BAO 2016 constraints on $\tau$ \citep{Planck-Collaboration:2016ab}. As expected, our early reionization model, RT5, is in tension with all constraints with reionization ending at $z_{\mathrm{reion}} \sim 8.5$, however we note that it is consistent with the PlanckTT+lowP+lensing+BAO 2015 constraints on the optical depth within $1 \sigma$ ($\tau = 0.066 \pm 0.013$, \citealt{Planck-Collaboration:2016aa}). We include this model to test the sensitivity of our results on the desired reionization history.

The difference between the volume- and mass-weighted neutral fraction becomes smaller as we increase the emissivity of stellar particles, as the increased production rate of photons boosts the ability of new sources to dominate the EoR. In our RT1.5 and RT2 models, the mass-weighted fraction levels out above zero, while the volume-weighted fractions of all simulations converge on zero, as the combination of an increased recombination rate and decreased stellar luminosity (lower emissivity) allows high-density regions to recombine on shorter time-scales.

The mass-weighted photoionization rates are significantly above the volume-weighted averages and constraints from \cite{Calverley:2011aa} and \cite{Wyithe:2011aa}, as ionization fronts emanate from high-density peaks and sweep through underdense regions much later. Due to the limited volume/statistics of our box, both the volume and mass-weighted photoionization rates show significant fluctuations at high-redshift. Towards the end of the EoR, the volume-weighted quantity smooths out as voids become ionized. The mass-weighted continues to show some fluctuation, especially in our high emissivity case (RT5), due to the balancing of photoionization and recombinations in high-density regions.

\subsection{Heating and Cooling Mechanisms}
Cooling is a crucial ingredient for galaxy formation. As gas crosses the virial sphere, a strong virialization shock results in the kinetic infall energy being thermalized and therefore heated to of order the virial temperature, which we define as:
\begin{equation}
T_{\mathrm{vir}} = \frac{1}{2} \frac{\mu m_{\mathrm{p}}}{k_{\mathrm{B}}} V_{\mathrm{c}}^{2},
\end{equation}
where
\begin{equation}
V_{\mathrm{c}}^{2} = \left( \frac{G \mathrm{M}_{\mathrm{vir}}}{R_{\mathrm{vir}}} \right)^{1/2}
\end{equation}
is the circular velocity of the halo at the virial radius. The gas then proceeds to form a hydrostatically supported atmosphere (hydrostatic equilibrium), which fills the halo with hot gas at densities several orders of magnitude below typical ISM densities. Therefore, the gas must cool and dissipate its thermal energy to enable further collapse towards the halo centre.

In the temperature range $10^{4}\ \mathrm{K} < T_{\mathrm{vir}} < 10^{6}\ \mathrm{K}$ or hot enough to collisionally excite hydrogen lines and partially or fully ionize the gas, atomic cooling dominates via a number of excitation and de-excitation processes, including recombination radiation and collisional excitation with subsequent decay. Above these temperatures, the gas is fully collisionally ionized and cools mainly via free-free emission (i.e.., Bremsstrahlung). Haloes whose virial temperatures are below the atomically cooling limit, i.e. $T_{\mathrm{vir}} \sim 8000\ \mathrm{K}$, are unable to collisionally ionize hydrogen and, therefore, are expected to be entirely neutral in absence of ionizing radiation. If gas is metal-free, the dominant cooling process in such haloes is via the rotational and/or vibrational lines of molecular hydrogen \citep{Abel:1997aa} and is crucial for the formation of the first stars. Finally, at very high redshifts, inverse Compton scattering of cosmic microwave background photons also provides an efficient cooling mechanism (Compton cooling).

With the exception of Compton cooling, all of the above processes require two particles; therefore, their efficiency scales with gas density. At high densities, if the gas has been pre-enriched with metals, fine-line transitions become an effective cooling mechanism, dominating in the ISM. Photoheating plays a crucial role in these cooling processes. The Lyman-Werner bands photons (11.2 to 13.6 eV), produced in copious amounts by young massive stars, excite H$_{2}$ molecules, resulting in their dissociation and thus suppressing star formation in mini-haloes. As this process is thought to sterilize mini haloes, we neglect detailed molecular chemistry here.

\subsection{Suppression Mechanisms}
\label{sec:suppression}
The significance of self-feedback by radiative heating/ionization is heavily debated in the literature, from its magnitude to whether it is a factor at all (see, for e.g \citealt{Shapiro:1994aa, Gnedin:2000aa, Shapiro:2004aa, Iliev:2005ab, Hoeft:2006aa, Wyithe:2006aa, Okamoto:2008aa, Kimm:2014aa, Ocvirk:2016aa}). Simulations of isolated disk galaxies show that radiative feedback acts primarily to prevent the adiabatic collapse of gas to higher (star-forming) densities, rather than push the gas past some temperature threshold deemed important for star formation \citep{Rosdahl:2015aa}. We repeat this investigation in a cosmological context, using our HD simulation to understand the impact of including radiative feedback in our RT models, focusing on the temperature and density distribution of gas in DM haloes.

\begin{figure*}
	\includegraphics[width=\textwidth]{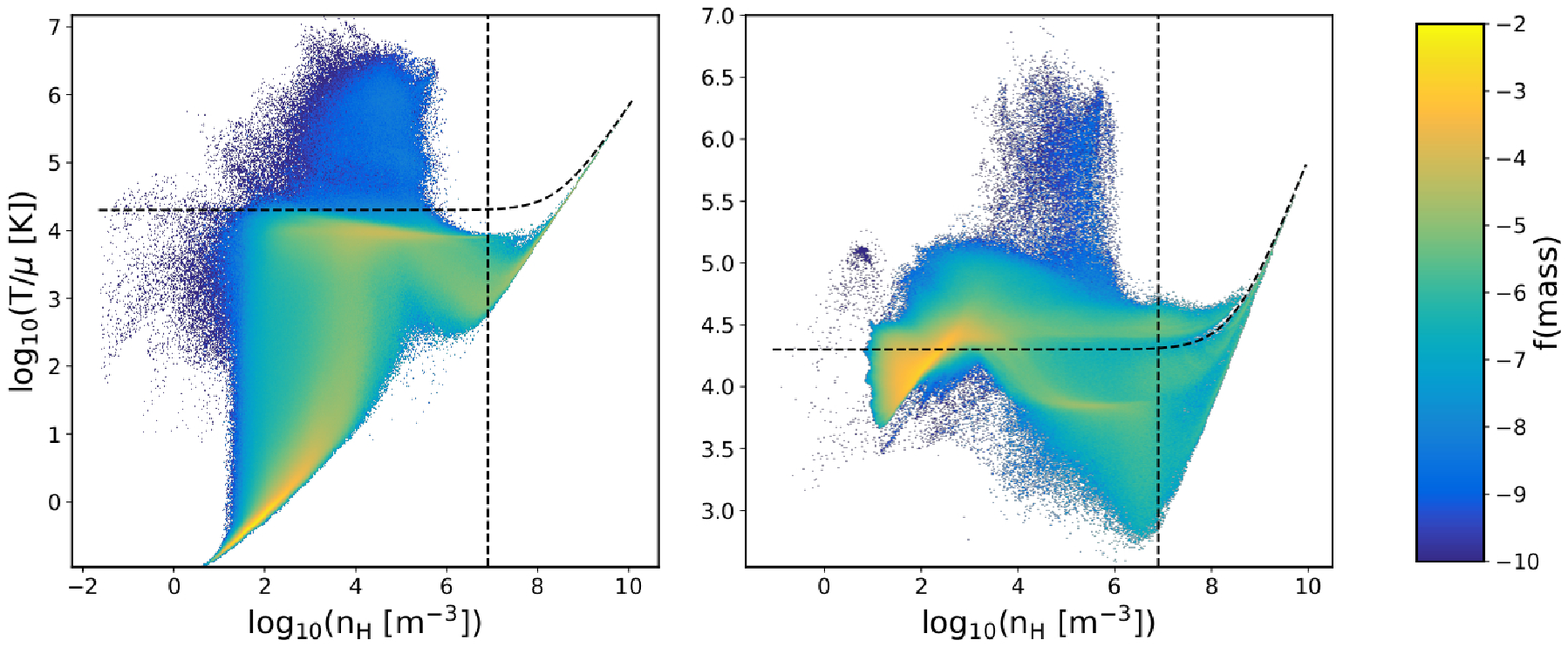}
    \caption{ Mass-weighted temperature-density phase diagrams for our HD and RT2 simulations (left/right respectively), for the full box at $z = 6.15$. The dashed lines represent the density and temperature criterion for star formation (vertical/horizontal), as discussed in Section~\ref{sec:calibration}, which bound the star-forming region of the plot (above/below the density/temperature criterion, respectively). The change in shape of the temperature criterion at $n_{\mathrm{H}} = n_{*}$ corresponds to the point where our polytropic temperature floor begins to dominate. In the SNe only case (HD), we see that most of the gas by fraction of mass remains on the adiabat $T/\mu \propto \rho^{\gamma - 1}$, where $\gamma$ is the ratio of specific hears. In the absence of a UV background, the gas is allowed to continue with adiabatic collapse into haloes, leading to run-away star and galaxy formation. In the RT2 case, the EoR raises the gas pressure of the IGM, preventing adiabatic collapse and raising the Jeans mass of galaxies. To collapse to higher densities, gas must cool along the visible atomic cooling branches at $T/\mu \sim 10^{4.6}\ \mathrm{K}$ and $T/\mu \sim 10^{3.8}\ \mathrm{K}$, which corresponds to the helium and hydrogen cooling peak cut-offs respectively. Once the gas reaches high densities, metal fine-line cooling becomes dominant and allows the gas to cool to $T/\mu \sim 10^{3}\ \mathrm{K}$ in both cases. At these high densities, gas is capable of self-shielding from ionizing UV, allowing star formation (see also Appendix~\ref{sec:cold_acc_streams}). }
\label{fig:phase_diag}
\end{figure*}

The evolution of gas in the temperature-density plane is shown in Fig.~\ref{fig:phase_diag} for our HD (left) and fiducial (right) models, where the vertical/horizontal dashed lines show the density and temperature criteria for star formation respectively. In the absence of radiative heating, the IGM remains cold with most of the gas occupying the adiabat $T/\mu \propto \rho^{\gamma-1}$, where $\gamma$ is the ratio of specific heats. Adiabatic collapse is therefore uninhibited; while in our fiducial model, photoheating makes the gas nearly isothermal at $T/\mu \sim 10^{4}\ \mathrm{K}$. The two visible branches in the right panel at densities $n_{\mathrm{H}} \gtrsim 10^{3} \mathrm{m}^{-3}$ result from atomic cooling peaks of hydrogen (lower) and helium (upper), which is highly efficient at these densities. This process allows the gas to collapse to the metal line cooling regime, reaching $T/\mu \sim 10^{3}\ \mathrm{K}/ \mu$ at densities $n_{\rm H} \gtrsim 10^{4} \mathrm{m}^{-3}$. Following the completion of the EoR (overlap stage), regions which are ionized later than average, predominately diffuse gas in the circumgalactic medium (CGM) or voids, remain hotter on average due to their longer cooling time ($t_{\mathrm{cool}} \sim 1/n_{\rm H}$) and additional local heating due to shocks from SNe and structure formation.

\begin{figure*}
	\includegraphics[width=\textwidth]{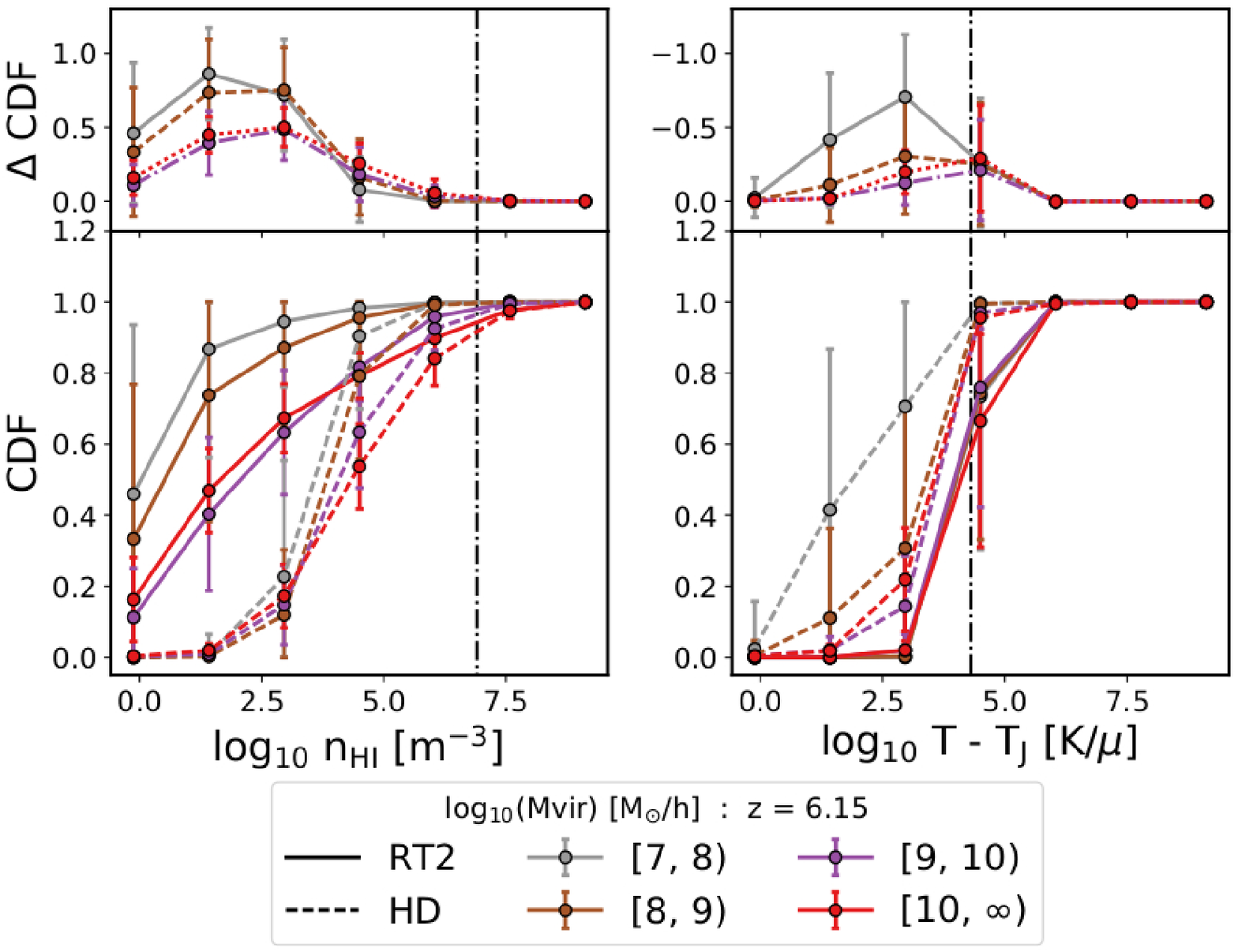}
\caption{ \emph{Bottom Left panel:} Mean cumulative distribution functions (CDF) of neutral hydrogen number density in logarithmic bins of halo mass for our fiducial RT2 model (solid) and SNe only (dashed) HD model at $z = 6.15$, where the error bars show the $1 \sigma$ standard deviation. The vertical dashed black (thin) line shows the density criterion for star formation, $n_{*}$. \emph{Bottom right panel:} Mean CDF of gas temperature minus the non-thermal Jeans pressure floor ($T_{\mathrm{J}}$. see Section~\ref{sec:calibration}), where the dot-dashed line shows the temperature criterion for star formation. The top panels for both plots show the difference between the RT2 and HD models. Radiation primarily acts to keep the gas in a diffuse state, preventing collapse to the halo centre and into galaxies. Low-mass haloes are most strongly affected due to their shallow potential wells, while haloes in the highest mass bin appear to show additional suppression over the next lower bin due to the amplification of SNe and radiative feedback \protect\citep{Pawlik:2009aa}. }
\label{fig:nHI_T_cdf_vir}
\end{figure*}

In both cases, feedback from SNe heats the gas to temperatures $2 \times 10^{4}\ \mathrm{K} < T/\mu < 10^{7}\ \mathrm{K}$, with noticeably less SNe heating when RT is included. As we will show below, this lack of SNe shock heating is due to the additional suppression of star formation by radiative feedback. As such events are already rare, any additional modulation of the star formation rate will have a significant impact on the SNe feedback energy budget. To determine whether radiation primarily acts to heat the gas above the star formation criteria or prevent collapse in the first place, we compute the cumulative distribution function (CDF) of densities and temperatures (left/right respectively) in haloes, binned by mass, as shown in Fig.~\ref{fig:nHI_T_cdf_vir}.

As gas infalls from the IGM onto haloes, it undergoes a strong virial shock, which heats it to of order the virial temperature. This heating leads to a hot diffuse atmosphere, which occupies the CGM. Irradiation can keep the gas in this diffuse state for prolonged periods of time, by extending the cooling time and thus preventing collapse. This behaviour is clearly demonstrated for our fiducial RT2 simulation in the bottom left panel of Fig.~\ref{fig:nHI_T_cdf_vir} (solid lines), when compared to our SNe only case (dashed lines), where we show the difference between both cases in the top panel. Radiative feedback significantly alters the statistical distribution of gas densities in haloes of all masses, with low-mass haloes most strongly affected (due to their shallow potential wells). This feedback leads to a reduction in gas density several orders of magnitude below $n_{*}$ (vertical dot-dashed line).

Interestingly, we find haloes in the highest mass bin are, on average, more suppressed in the high density regions than those in the $10^{9}\, M_{\odot} h^{-1} \leq \mathrm{M}_{\mathrm{vir}} < 10^{10}\ M_{\odot} h^{-1}$ bin when RT is included. This additional suppression is due to the amplification of SNe by radiative feedback and vice versa, as first discussed in \cite{Pawlik:2009aa}. As the gas is kept in a diffuse state, a Sedov blast wave can much more efficiently blow gas from the centre of the halo to the outer regions, where in turn it is more susceptible to photevaporation \citep{Pawlik:2009aa}. While this phenomenon is present in all haloes, it is most effective in high-mass haloes, where the frequency of SNe is higher.

In the right-hand panel, we compute the same quantity but for the gas temperature, where we have subtracted the Jeans pressure floor introduced in Section~\ref{sec:calibration}. The distribution of temperatures for all halo mass bins are roughly equal in our RT case, as photoheating raises the gas temperature to $\sim 8000\ \mathrm{K}$. In the absence of radiative heating, low-mass haloes contain significantly more cold gas due to their weak virial shocks and rare SNe events, bringing the mean gas temperature several orders of magnitude below the star forming criterion (vertical dot-dashed line). However, in the context of whether a cell can become star forming or not, we are only concerned with the properties of the ISM, where the typical gas density is $n_{\mathrm{H}} \sim 1\ \textrm{cm}^{-3}$. At these densities, cooling is extremely efficient as shown in Fig.~\ref{fig:phase_diag} and therefore not highly sensitive to the temperate criterion for star formation. Therefore, the primary mechanism by which radiation inhibits star formation is to prevent the otherwise efficient collapse from the IGM into galaxies.

\begin{figure*}
	\includegraphics[width=\textwidth]{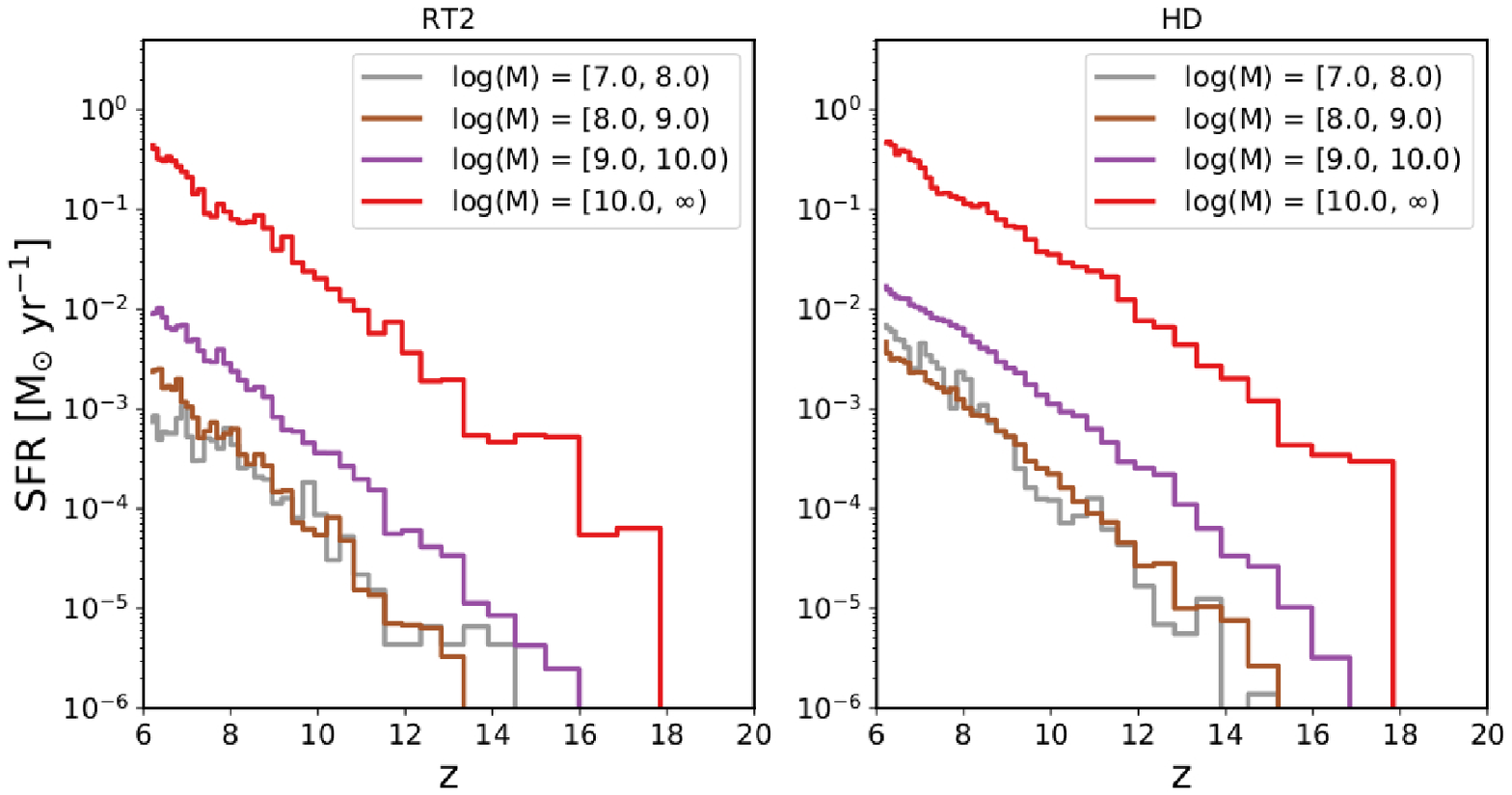}
\caption{ Star formation rates as a function of redshift in logarithmic bins of halo mass, in units of $M_{\odot} h^{-1}$, for our fiducial RT2 model (left) and HD (right). We compute the SFR by binning stellar particles in all haloes from $z=6$ back to the formation of the first star. Radiative feedback increasingly suppresses star formation towards low halo masses (by comparing left to right). Massive haloes are not strongly effected except at early times, when their progenitors are suppressed. By $z=6$, the SFR in the lowest mass halo bin has fallen by an order of magnitude with respect to the SNe only case, though does not cease entirely.  }
\label{fig:sfr_rt2_hd}
\end{figure*}

Fig.~\ref{fig:sfr_rt2_hd} shows the mean star formation rate (SFR) in logarithmic bins of halo mass (taken at $z = 6$) for our fiducial and HD models (left/right respectively), from high-mass (red) to low-mass (grey), as a function of redshift. High-mass haloes show very little suppression due to radiative feedback, except at early times when their progenitors are affected. As a result, large haloes typically undergo internal (inside-out) reionization, while low-mass haloes on average are reionized by distant sources (outside-in). This pattern is supported by simulations of the Local Group of galaxies, which find that low-mass satellites preferentially undergo external reionization due to the inside-out reionization imprinted by massive haloes \citep{Weinmann:2007aa, Ocvirk:2014aa}. While we do find that the SFR is increasingly suppressed towards lower halo masses, low-mass dwarfs are able to continue forming stars throughout the EoR, in contrast with recent works with lower numerical resolution \citep[e.g.][]{Ocvirk:2016aa}.

\begin{figure}
	\includegraphics[width=\columnwidth]{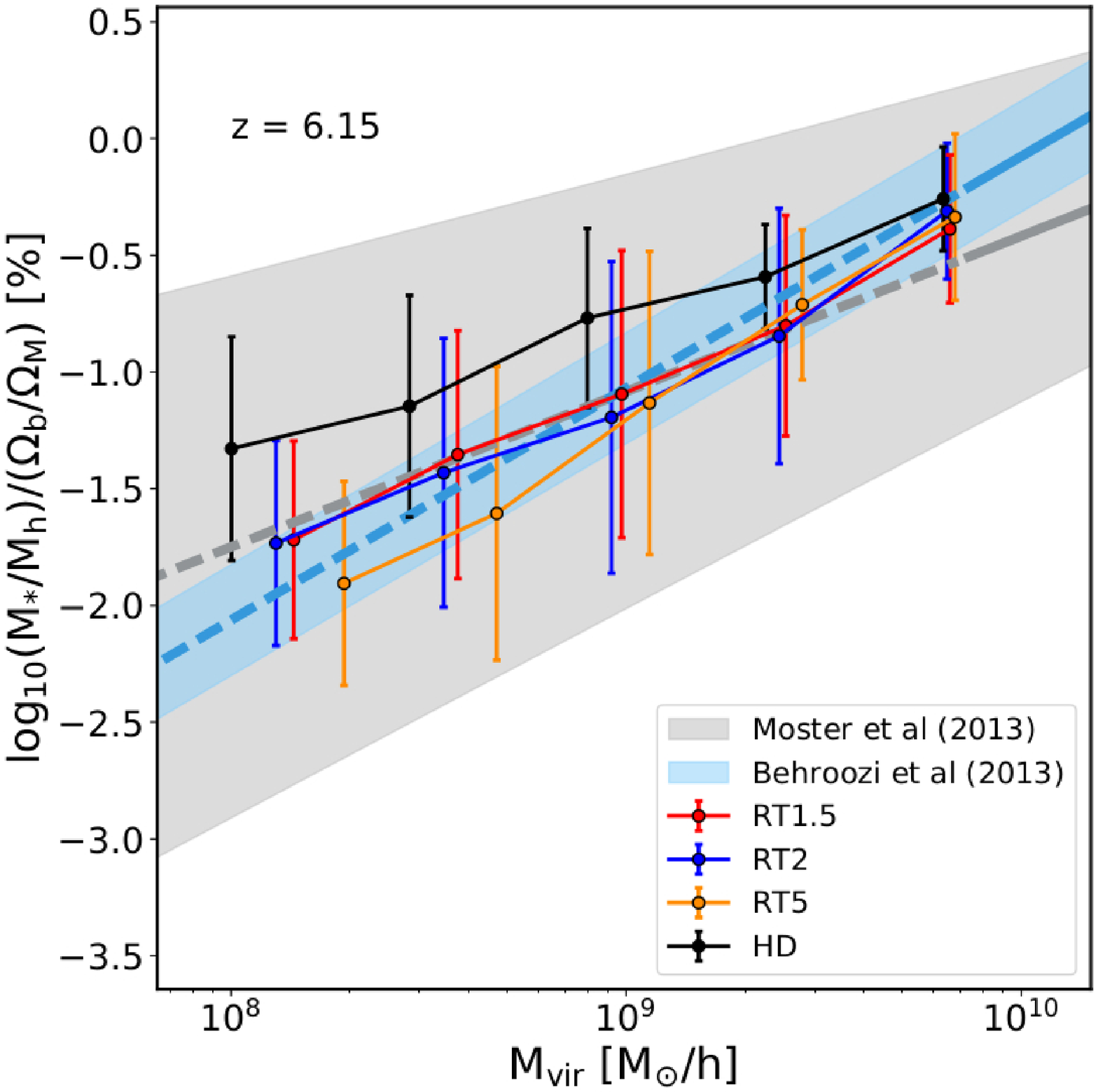}
\caption{ Stellar fraction of haloes (in units of the cosmic mean baryon fraction) for all simulations listed in Table~\ref{tab:sim_names} (points), compared with the models of \protect\cite{Moster:2013aa} and \protect\cite{Behroozi:2013aa}. Solid lines show the mean fit to the scatter, the error bars show the $1 \sigma$ deviation, and the dashed lines show the extrapolated values for both models. While haloes with masses $\mathrm{M}_{\mathrm{vir}} > 10^{8}\ M_{\odot} h^{-1}$ (i.e. $T_{\mathrm{vir}} \sim 10^{4}\ \mathrm{K}$) should not be strongly effected by UV heating, the legacy effect on their accumulated stellar mass, as their progenitors are suppressed, can be seen here when contrasted with the HD (SNe-only) case. Radiative feedback brings all RTX models into good agreement with the abundance matching model of \protect\cite{Behroozi:2013aa} by $\mathrm{M}_{\mathrm{vir}} \sim 8 \times 10^{8}\ M_{\odot} h^{-1}$, while all models fall well within the constraints of \protect\cite{Moster:2013aa}. }
\label{fig:stellar_abu_halo}
\end{figure}

Similarly in Fig.~\ref{fig:stellar_abu_halo}, we show the (log) stellar abundance as a percentage of the total halo mass, in units of the cosmic mean baryon fraction, at $z = 6.15$ for all models. We compare our results with the abundance matching models of \cite{Moster:2013aa} (grey) and \cite{Behroozi:2013aa} (blue), where the sold line/shaded region show the mean and $1 \sigma$ standard deviation respectively, and the dashed lines show the extrapolation below their minimum halo mass, $M_{\mathrm{vir}} \leq 10^{10}\ M_{\odot}$. We note that the \cite{Behroozi:2013aa} constraints are consistent with the observed galaxy stellar mass functions specific star formation rates and cosmic star formation rates from $z = 0$ to 8; however the \cite{Moster:2013aa} model only considers haloes from $z \sim 4$ to present, hence we extrapolate to our redshift range.

The reduction in stellar mass, as measured against our HD model (black), is proportional to the assumed sub-grid stellar emissivity, and thus the reionization history (red to orange, increasing). While we see a large reduction in the mean, only our RT5 model shows tension with the SNe-only case within the $1 \sigma$ error bars at low-masses. All of our models are well within the extrapolated (dashed lines) constraints of \cite{Moster:2013aa}, however radiative feedback is needed to bring our models into agreement with the extrapolated \cite{Behroozi:2013aa} constraints for intermediate halo masses. 

\subsection{The Halo Baryon Fraction}
\label{sec:bary_frac}
\begin{figure*}
\begin{center}
	\includegraphics[width=\textwidth]{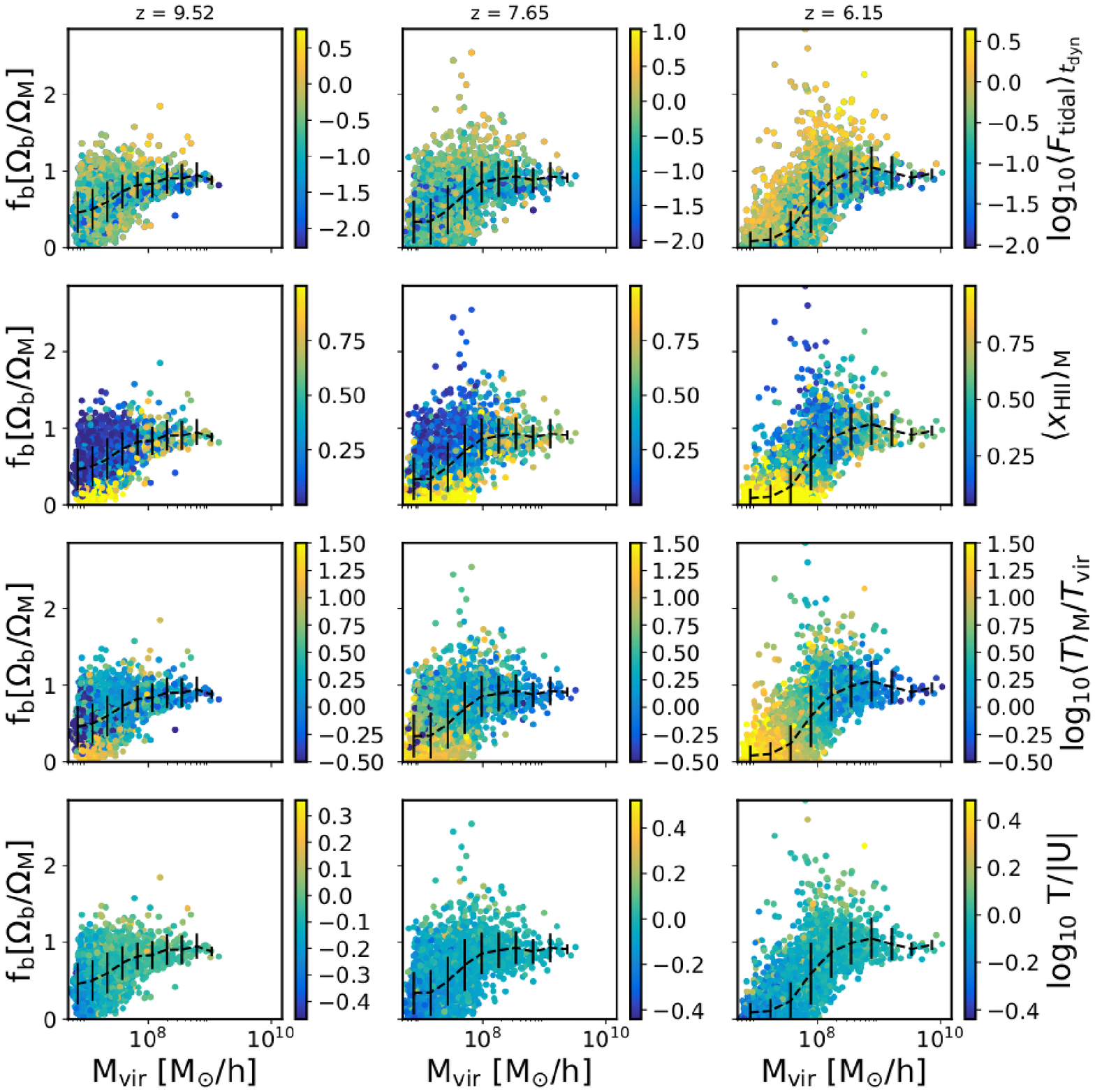}
	\caption{ Halo baryon fraction, in units of the cosmic mean, at redshifts 9.52, 7.65, and 6.15 for our RT2 simulation (columns; left to right). The colour bars show (top to bottom) the strongest tidal force exerted over the last dynamical time, mass-weighted hydrogen ionization fraction, mass-weighted gas temperature (in units of the virial temperature) and the virial ratio (ratio of kinetic to potential energies). The dashed lines show the mean baryon fraction in bins of logarithmic halo mass, where the error bars correspond to the standard deviation.}
\label{fig:fb_amr}
\end{center}
\end{figure*}

We compute $f_{\mathrm{b}}$ for each halo by summing the total mass of gas, stars and DM enclosed within the virial radius, $R_{\mathrm{vir}}$, as shown in Fig.~\ref{fig:fb_amr}. From left to right, we show the evolution over the redshift range $9.5 \lesssim z \lesssim 6$ for our fiducial model. The rows highlight the impact of four different quantities: the tidal force averaged over a dynamical time, $\langle F_{\mathrm{tidal}} \rangle_{t_{\mathrm{dyn}}}$, mass-weighted hydrogen ionization fraction, $\langle x_{\mathrm{HII}} \rangle_{\mathrm{M}}$, mass-weighted gas temperature in units of the virial temperature, $\langle T \rangle_{\mathrm{M}}/T_{\mathrm{vir}}$, and finally the (virial) ratio of kinetic to potential energies, $T/|U|$. We restrict our study to \emph{distinct} haloes only, as including sub-haloes may artificially inflate $f_{\mathrm{b}}$.

Prior to the EoR (left-hand column), $f_{\mathrm{b}}$ scatters about the cosmic mean for high-mass haloes, $\mathrm{M}_{\mathrm{vir}} > 10^{8}\ M_{\odot} h^{-1}$, and falls to roughly half this value at lower masses (black dashed line with error bars). In contrast, following the completion of reionization (centre and right-hand columns), low-mass haloes that are below the atomic cooling limit are quenched of their baryons, resulting in $f_{\mathrm{b}}$ falling to a few percent (or lower) of the total halo mass. In the high-mass regime, potential wells are deep enough to retain gas in the presence of an ionizing UV background, hence are not strongly effected. At all times, we find that a small percentage of the total halo population exhibit super-cosmic mean values of $f_{\mathrm{b}}$ \cite{Okamoto:2008aa} found that these were subhaloes at earlier times that have been tidally stripped from their hosts, leaving a dense baryon core. Such phenomena are expected in a hierarchical $\Lambda$CDM model of galaxy formation, and in extreme cases the entire galaxy can be disrupted leaving only a stellar stream (as is thought for the Sagittarius dwarf galaxy; \cite{Belokurov:2006aa}). To quantify this effect, we consider the strongest tidal force experienced from all neighbours (Fig.~\ref{fig:fb_amr}, top panel). The tidal force is defined as the ratio of the virial radius and the radius of the Hill sphere:
\begin{equation}
R_{\mathrm{Hill}} \simeq d \left( \frac{m}{3M} \right)^{1/3},
\end{equation}
where $d$ is the distance between the haloes, $m$ is the mass of the smallest halo and $M$ the mass of the largest halo. To determine whether the halo has had time to adjust, we average the tidal force over a single dynamical time $t_{\mathrm{dyn}} = (4 \pi G \rho)^{-1/2}$. 

\begin{figure*}
	\includegraphics[width=\textwidth]{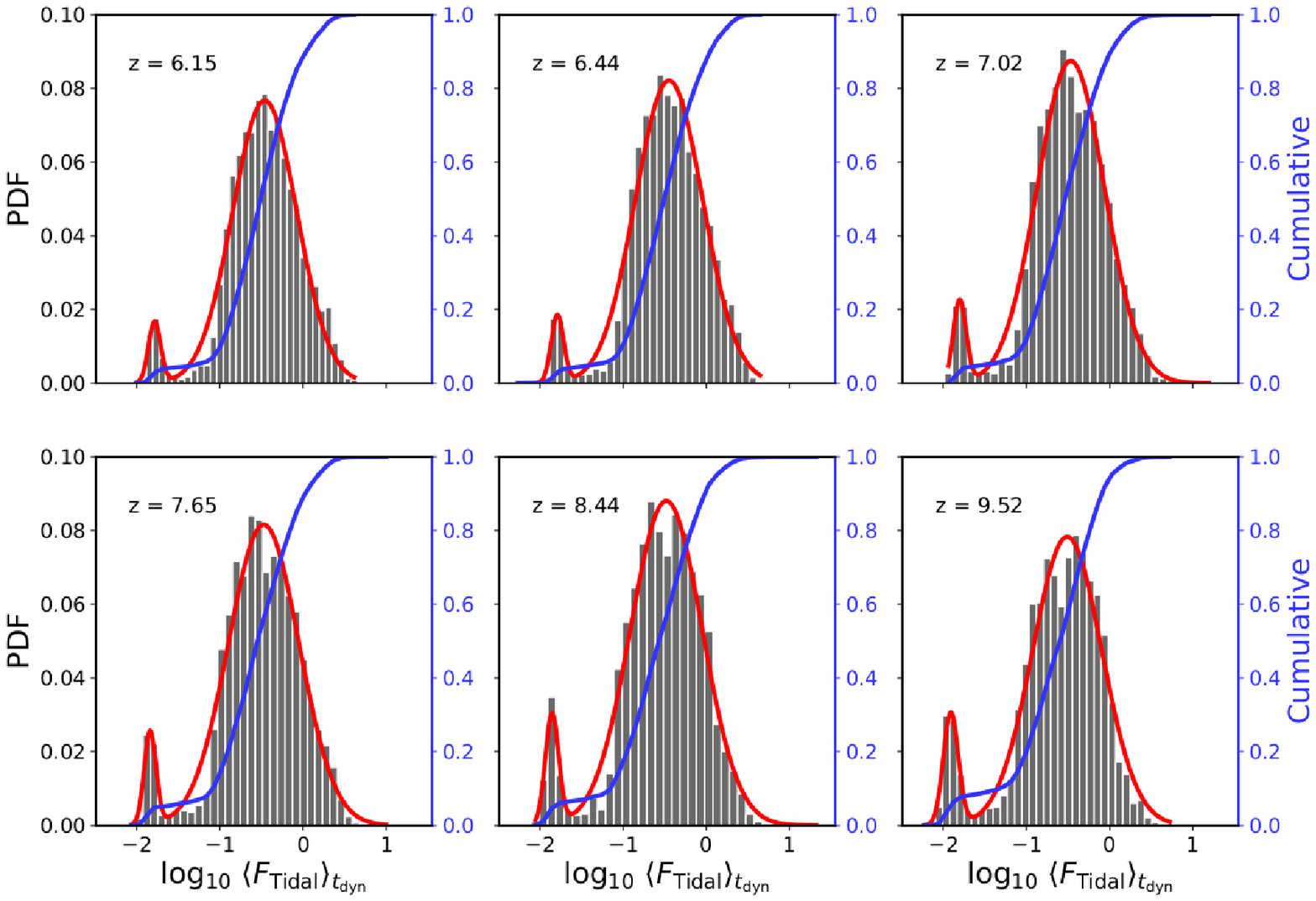}
\caption{ Probability distribution functions (PDFs) and Cumulative distribution functions (CDFs) for the tidal force averaged over a dynamical time, $\langle F_{\mathrm{tidal}} \rangle_{t_{\mathrm{dyn}}}$, from redshifts $6 \lesssim z \lesssim 9.5$. We fit a bimodal Gaussian distribution to each PDF (red solid lines). At all redshifts, approximately half of all haloes experience tidal forces less than or equal to the position of the larger peak, while the amplitude of the smaller peak slowly decreases over time as isolated galaxies eventually interact and (possibly) merge with more massive hosts. }
\label{fig:pdf_ftidal}
\end{figure*}

To better understand how the tidal force is distributed in the halo population, we show the probability (and cumulative) distribution functions (PDF and CDF respectively) for the tidal force in Fig.~\ref{fig:pdf_ftidal} between redshifts $6 \lesssim z \lesssim 9.5$ for our RT2 fiducial model. The PDF exhibits a strong bimodal distribution at all times, where the small peak characterises isolated haloes (near or in voids/underdense regions) that have not undergone any tidal interactions. Over time, the amplitude of the smaller peak slowly decreases over time as isolated galaxies eventually interact and (possibly) merge with more massive hosts. Approximately half of all haloes undergo strong tidal disruption, by which we mean the tidal force exceeds the mean of the larger peak, as shown by the CDF.

By comparing with the top row of Fig.~\ref{fig:fb_amr}, we find that strong tidal disruption (yellow points) leads to a deviation above/below the mean in all mass bins (black dashed line). This deviation is characteristic of a hierarchical model of galaxy formation, as haloes with large baryon fractions have stripped material from their neighbours, which are left baryon poor. In contrast, haloes that have not undergone significant disruption in their last dynamical time (blue points) fall on or close to the mean. The fraction of haloes in isolation that undergo very weak or no tidal forces decreases from $\sim 0.1$ at $z = 9.52$ to $< 0.5$ at $z = 6.15$, as demonstrated by the drop in the CDF where it intersects the lower peak. This leads to a reduction in the density of points which lie along the mean $f_{\mathrm{b}}$ in each mass bin, and therefore an increased scatter as shown by the error bars. High tidal forces typically lead to above average baryon fractions over time (yellow points; left to right), suggesting that tidal stripping may play a crucial role during the EoR; enabling haloes that might otherwise be devoid of gas to continue accretion. If this gas can cool/self-shield and condense towards the centre of the halo, this would enable the prolonged growth of any central galaxy, and continued star formation.

In the second row of Fig.~\ref{fig:fb_amr}, we show the impact of photoionization (indicated by the ionized fraction, second row of plots) on $f_{\mathrm{b}}$. As the EoR progresses (left to right), the fraction of dwarf haloes that are completely ionized increases, bringing the mean $f_{\mathrm{b}}$ down to a few percent of the total halo mass. In the centre column, reionization has only just finished (see Fig.~\ref{fig:reion_hist}), thus there has not been sufficient time for the photoevaporation of all haloes below the atomic cooling limit ($\mathrm{M}_{\mathrm{vir}} \sim 10^{8}\ M_{\odot} h^{-1}$). 

We find that immediately following the EoR, there are still dwarf haloes that contain cold gas (third row; dark blue points). If this gas is sufficiently dense, it may self-shield from the ionizing UV background, extending the time required to photoevaporate the halo fully. Between $z = 7.65$ and $z = 6.15$ (centre and right column), approximately $230\ \mathrm{Myr}$ passes, in which time the gas temperature rises and the photoevaporation of dwarfs takes place, bringing the mean $f_{\mathrm{b}}$ down from roughly half the cosmic mean to a few percent of the total halo mass.  At intermediate masses, between the gas poor and gas rich extremes, haloes can cool via atomic de-excitation and collisional processes, allowing the gas to recombine. 

Above these masses, $f_{\mathrm{b}}$ approaches the cosmic mean, despite large fractions of the gas being ionized (second row; yellow points at masses $M_{\mathrm{vir}} \gtrsim 10^{8}\ M_{\odot} h^{-1}$). This trend is due to the inside-out pattern of reionization imprinted by massive haloes, as discussed in Section~\ref{sec:suppression}. As their potential wells are large enough, photoheated gas is retained and remains as a hot, ionized, hydrostatically supported atmosphere in the CGM, where it may eventually cool and condense onto the central galaxy. The opposite is true for dwarf haloes, as these undergo outside-in reionization on average, hence potentially remain neutral until ionized by a distant source. Following their eventual photoevaporation, the only mechanism by which they are then able to accrete is via mergers and tidal stripping (as shown by the correlation between the yellow and dark blue points in the first and second rows of the third column, respectively).

Finally, in the bottom panels we show the virial ratio of haloes. At high-redshift, mergers are extremely common; therefore it is not always possible to discuss haloes as isolated systems (as is commonly the case when defining quantities based on the virial theorem). Haloes with excess kinetic energy (high virial ratio) have likely undergone strong interactions with a close neighbour, resulting in significant turbulence when baryons fall into the central regions \citep{Davis:2011aa}. This may play a crucial role in the cooling and collapse of baryons in high-redshift haloes, as a higher ratio of velocity dispersion to the sound speed can lead to increased fragmentation \citep{Clark:2011aa, Greif:2011aa}. We find that haloes with excess kinetic energy typically have super-cosmic mean baryon fractions, however only account for approximately 10 per cent of the total population at $z \sim 6$, in good agreement with \cite{Davis:2011aa}. Over time, the virial ratio of low-mass haloes falls as they relax and reach virial equilibrium.

\section{Artificial Neural Network}
\label{sec:ANN}

To explore the parameter space that influences the halo baryon fraction, we have developed an ANN, which we train using our fiducial simulation. The aim is to determine the minimum number of parameters required in order to accurately predict the simulation output, and more interestingly, make structured predictions based on variations in these parameters. Up to now, models fit the halo baryon fraction using a mass dependent fitting function \citep[e.g.][]{Henriques:2015aa, Lacey:2016aa}. However, as we have already shown in this work, the halo baryon fraction does not follow such a tight relation. Rather, a large amount of scatter is evident at all times, as one might expect, which can drastically alter the amount of available baryons for both galaxy and star formation. We demonstrate that by using ANNs and probing the impact of tidal forces, gas temperature, hydrogen ionization fraction, and virial ratio, one can accurately capture this scatter, construct a model that can reproduce these results, and drastically improve on existing methods.

Below, we briefly describe the architecture of our ANN and the choice of back-propagation algorithm used to train the model. We conclude this section by describing the adaptive-learning process we have implemented to improve the accuracy of our model in Section~\ref{sec:adaptive_learning_rate}.

\subsection{Architecture}

\begin{figure*}
	\includegraphics[width=\textwidth]{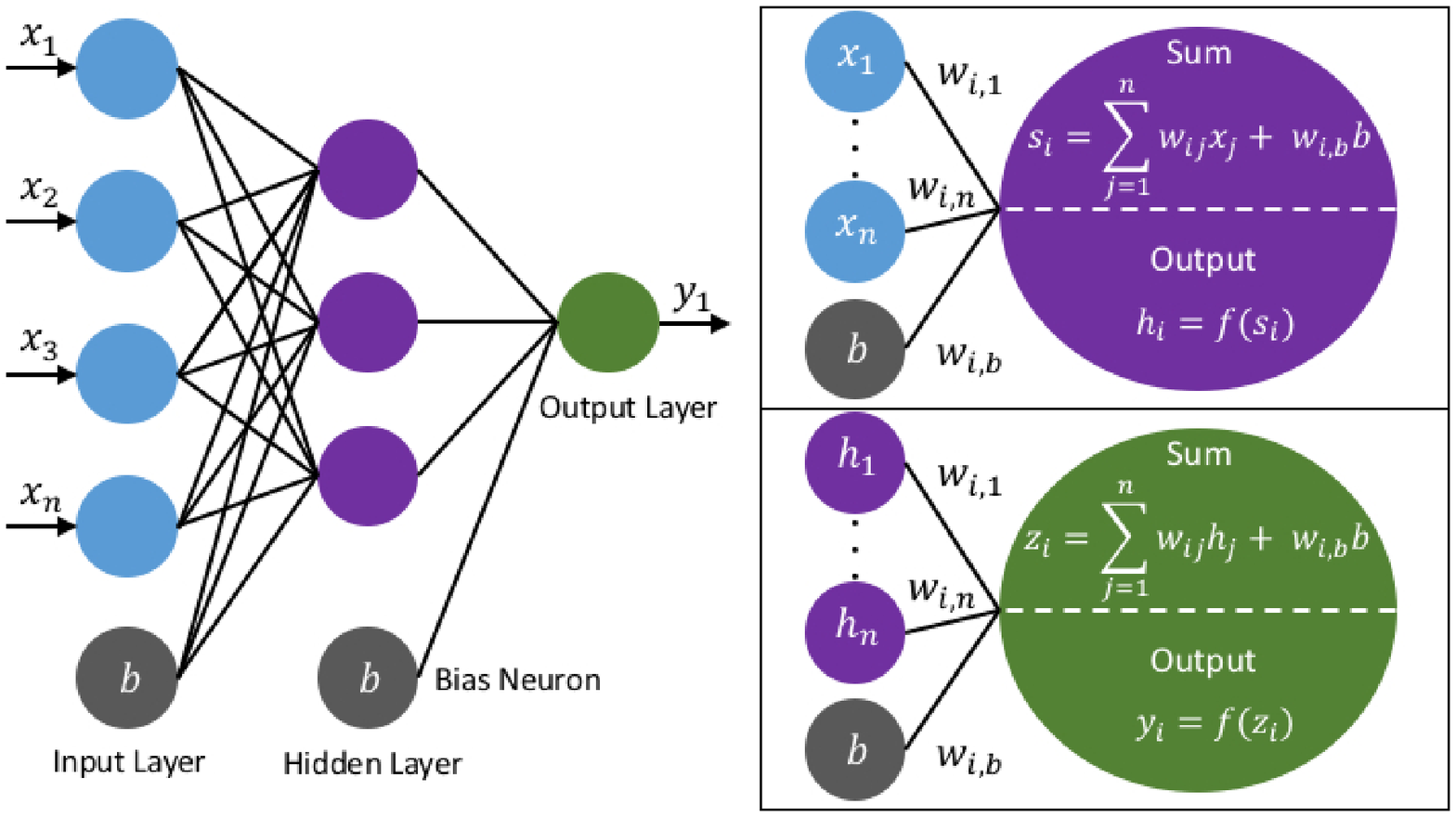}
    \caption{ Illustration of the architecture of our ANN, showing the input, hidden and output layer neurons. Each neuron is connected to all neurons in the previous layer with its own mathematical weight, $w$, over which the contributions are summed and applied to the transfer function as shown in the boxes on the right. We opt for non-linear sigmoid transfer functions for both the hidden and output layers. The grey neurons in the input and hidden layer denote bias neurons, which have their own connection weight as above, however have a constant output value of unity. By optimizing the weights of these neurons, the network can vary the transfer function along the input axis, meaning we do not require our input variable to be bound by the transfer function (however some scaling does improve the learning rate of the model). The ANN performs gradient descent at each neuron connection to minimise the cost function (Eqn.~\ref{eqn:ann_cost}) based on the target values supplied. We show three hidden neurons for illustrative purposes, however in practice we opt for a 4-45-1 topology (i.e 4 inputs, 45 hidden and 1 output). }
\label{fig:ann_arch}
\end{figure*}

Artificial neural networks are a branch of machine learning algorithms that mimic the natural neuron network in our brains through a series of mathematical weights, which are capable of learning any mathematical function to an accuracy proportional to the number of neurons in the network \citep{Cybenko1989, Hornik1989}. Many different ANN architectures exist, each suited to their own class of problems. At the simplest level, such a network consists of an input layer, one (or more) hidden layers, and a single output layer, each layer containing one or more neurons. Here, we develop an implicit, fully connected feed-forward network with a single hidden layer (shallow learning), as shown in Fig.~\ref{fig:ann_arch}, which is similar to that used in other works \citep[see, e.g.][]{Shimabukuro:2017aa}.

The input layer consists of one neuron for each input parameter, which simply accepts the input value and outputs it to each neuron in the next layer (often referred to as a latch neuron). The linear combination of the inputs is then calculated for each neuron, $s_{i}$, in the hidden layer as:

\begin{equation}
s_{i} = \sum\limits_{j=1}^n w_{ij}^{(1)} x_{j} + w_{i,b}b,
\label{eqn:input_hidden_linear}
\end{equation}
where neuron $j$ receives the input data $x_{j}$ and feeds it to the $i$-th neuron in the next layer with connection weight $w_{ij}$. The subscript $b$ denotes a bias neuron, which has a fixed output value of unity in our case. The superscript refers to the layer number (where 1 denotes the hidden layer and 2 denotes the output layer, introduced below). The purpose of such a neuron is to allow the network to vary its output along the input axis by modifying the weight $w_{i,b}$, such that the input values do not necessarily have to lie within a given range; whilst the traditional connection weights $w_{ij}$ control the steepness of the output. 

Each hidden neuron is then activated by applying the transfer/activation function, which can take a linear or non-linear form, depending on the problem and desired behaviour of the network. We opt for a non-linear transfer function, which is then applied such that our neuron output becomes $h_{i} = f(s_{i})$ and where the function $f(x)$ typically takes the form of a sigmoid function (as their derivatives are well known). The exact choice of transfer function again depends on the problem one desires to solve, as the training output must be bound by the output range of the neuron. For this work, we choose to use the hyperbolic tangent, however other choices may work equally as well. Each output neuron then sums the contributions from neurons in the hidden layer as above, such that:
\begin{equation}
z_{i} = \sum\limits_{j=1}^n w_{ij}^{(2)} h_{j} + w_{i,b}b
\label{eqn:hidden_output_linear}
\end{equation}
and again the non-linear transfer function is applied to give the final output $y_{i} = f(z_{i})$. While we have chosen to adopt both non-linear hidden and output layers, it should be noted that this is not strictly required, and one may prefer a non-linear transfer function for the hidden layer whilst simply taking the linear sum of weights for the output layer.

Finally, we may rewrite the above as a single equation, generalising the \emph{forward-propagation} of inputs through the network as:
\begin{align}
\begin{split}
s_{i}^{(\ell + 1)} &= \sum\limits_{j=1}^n w_{ij}^{(\ell)} x_{j}^{(\ell)} + w_{i,b}^{(\ell)}b^{(\ell)},
\\
y_{i}^{(\ell + 1)} &= f(s_{i}^{(\ell + 1)}),
\end{split}
\label{eqn:general_linear}
\end{align}
where $\ell$ refers to layers 1 or 2 as above. We hereafter reserve the notation $y_{i}$ (i.e without the super-script) for the output layer only.

\subsection{Back Propagation}

In this section, we discuss the backwards propagation algorithm used to train our ANN. For the network to quantify the accuracy of its predictions, we define the total cost function as:

\begin{equation}
E = \sum\limits_{n=1}^N \left[ \frac{1}{2} \sum\limits_{i=1}^m \left( y_{i,n} - \hat{y}_{i,n} \right)^{2} \right]
\label{eqn:ann_cost}
\end{equation}
where $N$ is the total number of training samples, $y_{i,n}$ denotes the neuron output for this training pass, and $\hat{y}_{i,n}$ the target value. The network performs gradient descent at each neuron connection to minimise this cost function by updating the internal system of weights, such that:

\begin{align}
\begin{split}
\Delta w_{ij}^{(\ell)} &= \eta \frac{\partial E}{\partial w_{ij}^{(\ell)}} + \xi \Delta \tilde{w}_{ij}^{(\ell)};
\\
\frac{\partial E}{\partial w_{ij}^{(\ell)}} &= \sum\limits_{n=1}^N \frac{\partial E_{n}}{\partial w_{ij}^{(\ell)}},
\label{eqn:ann_delta_weight}
\end{split}
\end{align}
where $\eta$ and $\xi$ are defined as the learning rate and momentum respectively and $\Delta \tilde{w}_{ij}^{(\ell)}$ is the change in the connection weight on the last training pass. These parameters are used to control how quickly the network learns and performs its gradient descent, which we allow to vary during the learning process and discuss in greater detail in the following section.

The derivative of the cost function with respect to the weights between the input and hidden layer can be calculated as:
\begin{align}
\begin{split}
\left( \frac{\partial E_{n}}{\partial w_{ij}^{(\ell)}} \right)^{(\ell)} &= \frac{\partial E_{n}}{\partial s_{i}^{(\ell)}} \frac{\partial E_{i}}{\partial w_{ij}^{(\ell)}}
\\
 &= \left[ \frac{\partial E_{n}}{\partial y_{i}^{(\ell)}} \frac{\partial y_{i}^{(\ell)}}{\partial s_{i}^{(\ell)}} \right] x_{j}^{(\ell)}
 \\
 &= \left[ \delta^{(\ell)} f' \left(s_{i}^{(\ell)}\right) \right] x_{j}^{(\ell)}.
\end{split}
\label{eqn:ann_error_prop}
\end{align}
Here, we have introduced the \emph{error propagation} term $\delta^{(\ell)}$, where for the first layer:

\begin{equation}
\delta^{(1)} \equiv \sum\limits_{n=1}^m \left( y_{i,n} - \hat{y}_{i,n} \right) w_{ij}^{(2)}
\label{eqn:err_input_hidden}
\end{equation}
and $m$ is defined as the number of neurons in the next layer. Note that in getting from equation~(\ref{eqn:ann_delta_weight}) to equations~(\ref{eqn:ann_error_prop}) and~(\ref{eqn:err_input_hidden}) we have used the chain rule for the derivative, as $E$ depends solely on the activated hidden neurons.

For the output layer, the error-propagation term is simply the difference between our network output and the target value multiplied by the connection weight:

\begin{equation}
\delta^{(2)} \equiv \left( y_{i,n} - \hat{y}_{i,n} \right)
\label{eqn:err_hidden_output}
\end{equation} 
A complete iteration to train the network therefore requires carrying out the following steps:

\begin{enumerate}
\item Initialise the network with random connection weights
\item Compute the outputs using equation~(\ref{eqn:general_linear}) (Forward-propagation)
\item Compute the output gradients using equations~(\ref{eqn:ann_error_prop}) and~(\ref{eqn:err_hidden_output})
\item Compute the hidden layer gradients using equations~(\ref{eqn:ann_error_prop}) and~(\ref{eqn:err_input_hidden})
\item Update the set of internal weights using equation~(\ref{eqn:ann_delta_weight}) (Back-propagation)
\end{enumerate}
Once the network has been trained, the connection weights are exported and re-imported as required.

\subsection{Adaptive Learning Rate}
\label{sec:adaptive_learning_rate}

Generally, there is no standardised way of determining the appropriate learning rate and momentum of a neural network. One tends to use trial and error to determine the best choice of $\eta$ and $\xi$, which varies based on the number of neurons and layers used. Typically, we want the initial learning rate to be fast, such that the network quickly determines a set of internal weights which roughly approximate the target function, without requiring a large number of training passes. However, this requirement could limit the final accuracy of the model, as large steps are taken in the gradient descent process thus making it difficult to converge on the global minimum. One solution to this problem is to allow the network to adaptively vary its own learning rate as the total root-mean-square (RMS) error falls below an acceptable threshold.

\begin{figure}
	\includegraphics[width=\columnwidth]{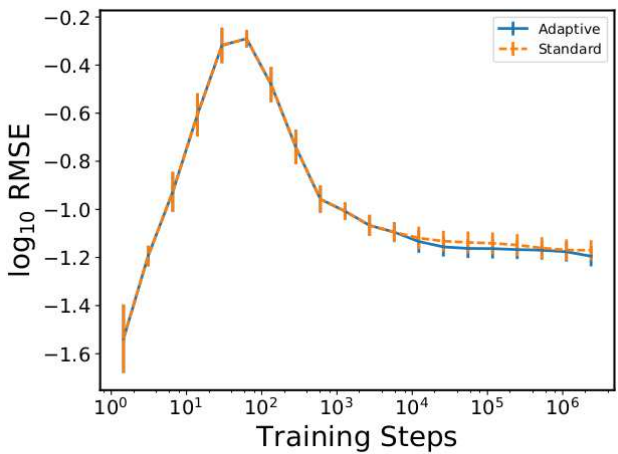}
    \caption{ RMS error of the ANN, for the adaptive (varying; blue) and standard (fixed; orange) learning rates. Both models produce identical errors for the first $10^{3}$ iterations, after which the adaptive method begins to reduce the net learning rate $\eta$ and momentum $\xi$, allowing the gradient descent to get closer to a global minimum. While the difference appears small, this mostly corresponds to improved accuracy at capturing rare outliers in the data. }
\label{fig:ann_rms}
\end{figure}

We opt for a simple adaptive learning implementation, whereby the parameters $\eta$ and $\xi$ are initially set to 0.15 and 0.25, and subsequently allowed to fall by a factor of two and four when the RMS error is measured to be consistently below some threshold. In Fig.~\ref{fig:ann_rms} we show the resulting performance of the network with/without adaptive learning, where the error bars show the standard deviation in the RMS for each bin. The input data consist of a few thousand haloes, which the network makes several passes over in a random order such that the total number of iterations exceeds $10^{6}$. As expected, the initial performance is identical in both cases, however the adaptive learning rate leads to a subtle reduction in the mean RMS after roughly $10^{4}$ training passes. While the effect appears small, the adaptive learning rate allows the network to achieve a higher accuracy with sparse data sets and thus greater accuracy when capturing rare outliers in the data

\subsection{Constraining $M_{\mathrm{c}}$ with Artificial Neural Networks}
\label{sec:constrain_Mc}

In line with recent studies \citep{Noh:2014aa, Okamoto:2008aa, Hoeft:2006aa}, we quantify the scale at which the baryon fraction transitions from poor to rich using the \cite{Gnedin:2000aa} fitting formula:
\begin{equation}
\label{eqn:gnedin}
f_{\mathrm{b}}(M,z) = \langle f_{\mathrm{b}}  \rangle \left\lbrace 1 + \left(2^{\alpha / 3} - 1\right) \left[ \frac{M}{M_{\mathrm{c}}(z)}  \right]^{-\alpha} \right\rbrace^{-3 / \alpha},
\end{equation}
where $\langle f_{\mathrm{b}}  \rangle$ denotes the cosmic mean baryon fraction, $M$ is the halo mass, $M_{\mathrm{c}}(z)$ is the redshift dependent characteristic mass scale defined earlier, and $\alpha$ is an exponent which controls the steepness of the transition between the two extremes. The characteristic mass scale represents a halo whose baryon fraction is half the cosmic mean value, as determined from hydrodynamical simulations and necessary to fit the baryon fraction in semi-analytical models of galaxy formation (such as the Munich and Durham semi-analytical models; i.e. {\sc L-galaxies} and {\sc galform}, respectively; \citealt{Henriques:2015aa, Lacey:2016aa}). While the exponent need not be constant, a value $\alpha = 2$ is found to give good agreement with our results.

It was recently found by \cite{Aubert:2015aa} that $f_{\mathrm{b}}$ is independent of reionization history at fixed \emph{volume-weighted} ionized fraction. This is significant, as we have departed from thinking about how $f_{\mathrm{b}}$ varies in time (redshift), to instead considering how it varies based on the physical evolution of haloes. This paradigm shift is our primary motivation in developing an ANN to understand, and predict this quantity based solely on time-independent physical quantities.

\begin{figure*}
	\includegraphics[width=\textwidth]{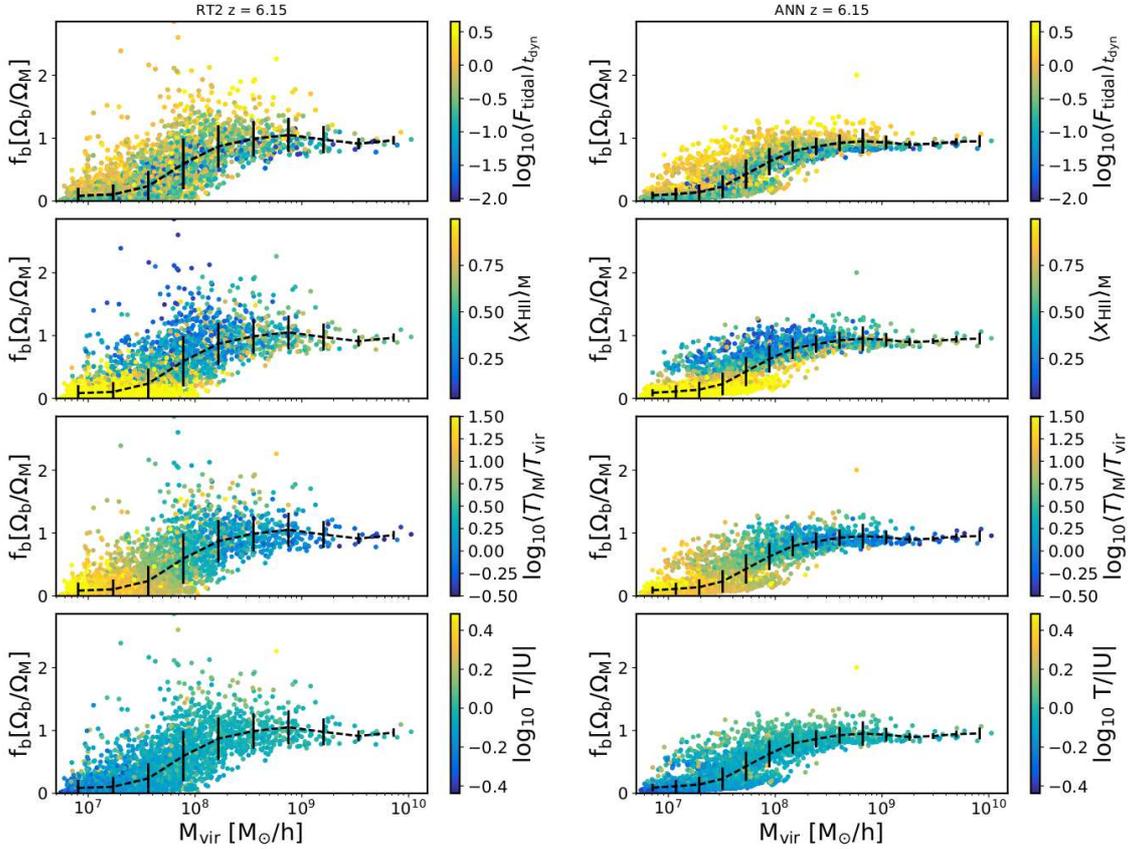}
    \caption{ Same as Fig.~\ref{fig:fb_amr}, but for our fiducial model (left) and ANN predictions (right) at $z = 6.15$. We use the existing halo catalogue data, as it is presented in the left-hand column, for our network predictions. The ANN is trained using only our fiducial simulation, and is then tested against the remaining RT1.5 and RT5 models with $M_{\mathrm{c}}$ predictions shown in Fig.~\ref{fig:Mc_z_xHII_tidal_cutoff} (shaded is simulated, error bar points are our ANN predictions). The model shows extremely good agreement for all quantities shown, however struggles to match the most extreme cases (only a few percent of haloes) as we train over all redshifts where the degree of scatter varies over time, therefore the network smooths over the rarest objects. }
\label{fig:fb_amr_ANN}
\end{figure*}

We train the model with our fiducial simulation only, using the four quantities shown in Fig.~\ref{fig:fb_amr} as our input parameters. The training data consists of haloes from eight snapshots between $z \sim 12$ and $z \sim 6$, which we write out randomly over several thousand passes such that the total number of training samples exceeds $10^{6}$. Once trained, we export the internal system of connection weights, which is all that is required to calibrate the model for future use. We use this model to predict $f_{\mathrm{b}}$ based on our simulated halo catalogues, to test the accuracy of the network, and present our results in Fig.~\ref{fig:fb_amr_ANN} for our fiducial simulation (left column) and ANN (right column) at $z = 6.15$.

Our model reproduces the simulated data well for all quantities, matching the trends for each quantity except for the rarest objects. These represent only a few percent of the total halo population, and require additional physics in our training data to fully capture. While the scatter is reduced in our predictions, as shown by the black dashed fit and error bars, it is a remarkable improvement over existing constraints which are mass and redshift dependent only. The network captures many interesting properties from our simulated haloes, such as the correlation between high tidal forces and large neutral fractions at low-masses and the presence of high tidal forces above and below the mean, with relaxed haloes following the best fit curve. The overall accuracy could be further improved by including more input parameters, to increase the number of degrees of freedom in the model, however we restrict our work to a limited sample as we aim to improve usability in SAMs of galaxy formation; which may not be able to quantify a large range of inputs accurately.

\begin{figure*}
	\includegraphics[width=\textwidth]{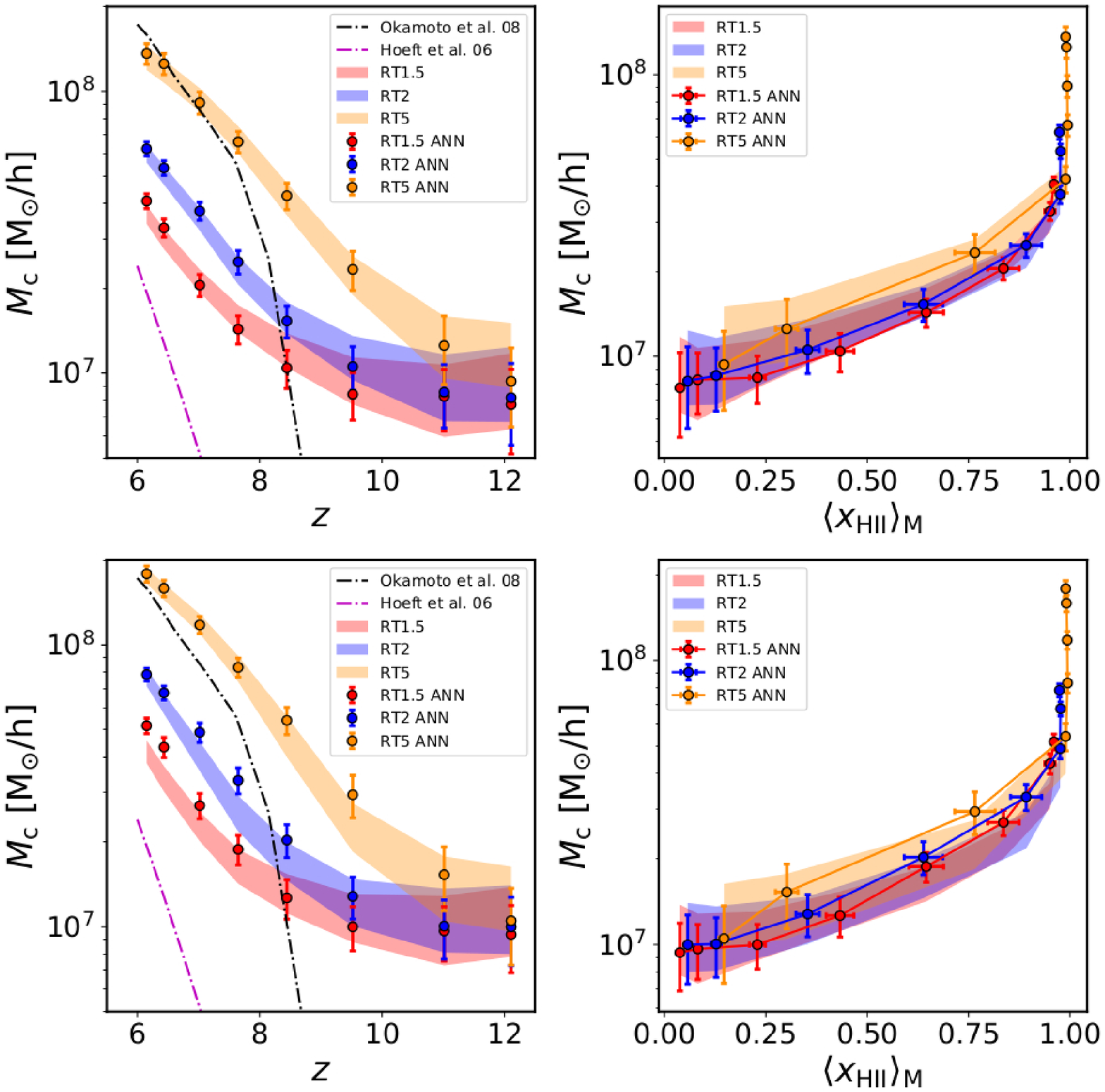}
\caption{ \emph{Top left panel:} Evolution of the characteristic mass scale, $M_{\mathrm{c}}$, as a function of redshift for all RTX models listed in Table~\ref{tab:sim_names} (shaded), and our ANN predictions (points). The error bars correspond to the $5 \sigma$ confidence level. The dot-dashed lines show the fits from \protect\cite{Okamoto:2008aa} (black) and \protect\cite{Hoeft:2006aa} (magenta), who assume a quasi-homogenoeus UV background using the models of \protect\cite{Haardt:2001aa} and \protect\cite{Haardt:1996aa} respectively. \emph{Top right panel:} Evolution of $M_{\mathrm{c}}$ as a function of mass-weighted hydrogen ionization fraction, showing that all model are consistent at fixed points in their reionization histories. Note, we do not include the results of \protect\cite{Okamoto:2008aa} and \protect\cite{Hoeft:2006aa} as their models transition from a neutral to fully-ionized Universe instantaneously (step function). \emph{Bottom panels:} Same as above, but with a cut-off in the tidal force below the position of the highest peak in the redshift dependent PDF (see Fig.~\ref{fig:pdf_ftidal}) applied to the simulated data. Note, for our ANN predictions we re-sample the tidal force from the PDF below the same cut-off, but for all haloes as a toy model to test the predictive power of the model. The ANN is trained using simulated data from our fiducial simulation only, and then used to predict $M_{\mathrm{c}}$ for RT1.5 and RT5. While the ANN predictions for RT1.5 in our tidal cut-ff model (bottom left panel) are slightly high, overall the network matches the simulated data extremely well. }
\label{fig:Mc_z_xHII_tidal_cutoff}
\end{figure*}

To probe accuracy over a range of redshifts, we compute $M_{\mathrm{c}}(z)$ by non-linear least squares fitting to Eqn.~\ref{eqn:gnedin} for a range of snapshots between $12 \leq z \leq 6$. The exponent, $\alpha$, is left as a free parameter, however a value of two is found to give good agreement at all times. This is done for all simulations listed in Table~\ref{tab:sim_names} (shaded) and the corresponding ANN predictions at the same redshifts (points), as shown in the top left panel of Fig.~\ref{fig:Mc_z_xHII_tidal_cutoff}. Our fiducial simulation falls significantly below the predictions of \cite{Okamoto:2008aa} (black dot-dashed line) shortly after their instantaneous reionization at $z_{\mathrm{reion}} = 9$, and above at earlier times. This is due to the inhomogeneous nature of reionization in our simulations, as \ion{H}{ii} regions begin to grow around the first stars which form as early as $z \sim 18$ (see Fig.~\ref{fig:sfr_rt2_hd}). While this leads to an increased suppression of the halo baryon fraction prior to the EoR, as $M_{\mathrm{c}}(z)$ is significantly larger, we find that haloes are not as strongly suppressed as previously thought following its completion. In fact, only our early reionization simulation, in which haloes are suppressed much earlier in their lifetimes, gives good agreement by $z \sim 6$. We also show the predictions from \cite{Hoeft:2006aa}, who use a similar homogeneous UV background model as discussed above, but trigger instantaneous reionization at $z_{\mathrm{reion}} = 6$; hence their constraints initially lag behind before converging at the present day ($z = 0$).

The ANN, trained on our fiducial haloes only, shows extremely good agreement at all times, with the uncertainties becoming smaller with redshift in line with our simulations. As $M_{\mathrm{c}}$ sets the mass scale where $f_{\mathrm{b}}$ transitions from a few percent to the cosmic mean, this shows that our model can accurately predict the shape of this relation independent of the underlying reionization history, with the introduction of physical scatter in our $f_{\mathrm{b}}$ predictions. This represents a significant improvement over existing constraints on $M_{\mathrm{c}}(z)$, which are model, mass, and time dependent. To demonstrate that our findings are independent of redshift, we show the same quantity but as a function of \emph{mass-weighted} hydrogen ionized fraction in the top right panel of Fig.~\ref{fig:Mc_z_xHII_tidal_cutoff}. We find good agreement with \cite{Aubert:2015aa}, however note that they compare the halo baryon fraction as a function of \emph{volume-weighted} ionized fraction. When investigating the internal properties of haloes, we are primarily concerned with the mass structure over the total volume, hence we opt for the mass-weighted quantity. This demonstrates that the Jeans filtering of haloes during the EoR is a \emph{local} effect, and not globally time dependent. Therefore, redshift dependent constraints on $M_{\mathrm{c}}$ are not well motivated physically, and miss much of the scatter in the galaxy population during this time.

In the bottom two panels of Fig.~\ref{fig:Mc_z_xHII_tidal_cutoff}, we show the same as above but with a tidal cut-off, whereby we exclude all haloes where $\langle F_{\mathrm{tidal}} \rangle_{t_{\mathrm{dyn}}}$ exceeds the value of the highest peak in the PDF at each redshift (Fig.~\ref{fig:pdf_ftidal}). \cite{Okamoto:2008aa} were similarly motived to remove haloes which had undergone recent merger activity to reduce the super-cosmic mean scatter in their simulations. As demonstrated in Fig.~\ref{fig:fb_amr}, the net effect of tidal stripping is to raise the baryon fraction above the halo average over time (top row, left to right: high density of yellow points above the dashed mean). Therefore, we find that excluding such objects leads to a significant rise in $M_{\mathrm{c}}$ for all models. To test the sensitivity of our ANN to such a scenario, we create a mock training sample whereby we re-sample the tidal forces exerted on all haloes from the PDFs shown in Fig.~\ref{fig:pdf_ftidal}, but below the large peak as done above. The model correctly predicts a rise in $M_{\mathrm{c}}$ for all simulations, with minor tension shown for RT1.5. This tension is likely due to the random re-sampling of the PDF, however the model still accurately predicts the distribution of all other quantities (see Appendix \ref{sec:low_tidal_force}, Fig.~\ref{fig:fb_amr_tidal_cutoff}). In the bottom right panel, we find that all models are still consistent with each other as a function of mass-weighted ionized fraction, suggesting that photoionization/heating are the dominant processes in controlling $f_{\mathrm{b}}$.

\section{Discussion}
\label{sec:discussion}

The Jeans-mass filtering of low-mass galactic haloes embedded within \ion{H}{ii} regions has long been proposed as an efficient mechanism to regulate star formation, by preventing the accretion of baryons from the photoheated IGM \citep[e.g][]{Shapiro:1994aa, Quinn:1996aa, Gnedin:2000aa}. Haloes that achieve a high stellar emissivity will reach a much higher ionized fraction earlier than those with low star-forming efficiencies, which in turn will lead to them undergoing Jeans-mass filtering earlier. This effect compensates for the difference in emissivity of haloes in the suppressible regime, leading to a `self-regulating' reionization \citep{Iliev:2007ab}, and could help explain the discrepancy between the observed number density of Milky Way dwarf satellites and those predicted by $N$-body simulations that assume a $\Lambda$CDM cosmology. However, previous studies on this subject have failed to intricately resolve the internal structure of galactic haloes during reionization, potentially overestimating the efficiency of radiative feedback on star formation.

Semi-analytical models predict that the modulation of star formation due to Jeans-mass filtering at the reionization redshift should be observable in both the cosmic star formation rate of low-mass galaxies and the faint end of the UV luminosity function \citep[e.g][]{Barkana:2000aa}. Recent RHD simulations of the EoR have also drawn this conclusion, showing that the comoving star formation rate of the lowest mass haloes, below masses where atomic cooling is possible, are significantly suppressed and in some cases can even be quenched entirely \citep[e.g][]{Ocvirk:2016aa}. Large scale $N$-body + RT simulations, such as those by \cite{Iliev:2007ab} and more recently \cite{Dixon:2016aa}, which probe the global geometry and statistics of reionization, incorporate these assumptions into their models as sub-grid physics in order to determine which sources drive reionization.

High-resolution simulations of isolated disk galaxies, such as those carried out by \cite{Rosdahl:2015aa}, have shown that radiative feedback primarily acts to prevent the collapse of gas and thus the formation of star-forming clumps, rather than destroy those that have already formed.  \cite{Rosdahl:2015aa} conclude that the reduction in star formation due to radiative feedback ranges from a factor of 4 in low-mass galaxies to roughly 0.1 for the most massive, similar to that of traditional thermal dump SNe feedback models. This reduction is much weaker than recently predicted by large-scale RHD simulations of the EoR and assumed in the sub-grid models of $N$-body + RT simulations, such as those discussed above.

One of the primary aims of this paper was to determine the statistical effect of radiative feedback on galaxies in a relatively small cosmological volume. In Section \ref{sec:suppression}, we carried out a statistical analysis of the distribution of gas densities and temperatures within haloes and concluded that while radiative feedback can heat the gas above our temperature threshold for star formation, it did not have a significant impact on galaxies, as gas at typical ISM densities can already cool efficiently. The CDF of the neutral gas of haloes exposed to ionizing radiation in our simulations show a large reduction in the fraction of dense gas at all masses with respect to the SNe-only case, with low masses most strongly effected. These lower on average densities lead to a reduction in the star formation rate and stellar abundance of haloes; however, low-mass haloes are not completely quenched. As such, we conclude that radiative feedback is much weaker than previously thought, in good agreement with studies of isolated disks \citep[e.g][]{Rosdahl:2015aa}.

To understand why the lowest-mass galaxies are able to continue forming stars in our simulations, we investigated how various physical effects impacted the baryon fraction of haloes, $f_{\mathrm{b}}$. These included tidal forces, photoionization, heating, and the degree to which haloes are virialised at high redshifts. The Jeans-mass filtering and subsequent photoevaporation of the lowest-mass haloes leads to a reduction in $f_{\mathrm{b}}$ to a few percent of the total halo mass, while intermediate to high-mass haloes are able to retain a significant fraction of their baryons. This effect becomes more exaggerated towards to end of reionization, as photoevaporation occurs on time-scales of a few hundred million years \citep{Shapiro:2004aa, Iliev:2005ab}. The highest-mass haloes remain at the cosmic mean baryon fraction, despite radiation preventing collapse and keeping gas at low densities, as previously demonstrated by the CDF of neutral hydrogen. Their large potential wells can retain photoheated gas, remaining as an ionized, hot, hydrostatically supported atmosphere. Tidal stripping leads to a large scatter in $f_{\mathrm{b}}$ above the median, allowing the lowest-mass haloes to accumulate significant fractions of neutral gas. If this gas can cool and collapse to high densities, it may become self-shielding from ionizing UV radiation and thus allow prolonged star formation. The ability for the gas to become self-shielding requires extremely high-resolution simulations, which resolve gas at extremely high overdensities ($\rho / \langle \rho \rangle \gg 10^{5}$). At these high densities, self-shielding could also enable the formation of cold accretion streams, which allow massive galaxies to continue to infall baryons throughout reionization (see Appendix~\ref{sec:cold_acc_streams} and \citealt{Rosdahl:2012aa}). Lower resolution simulations which do not resolve these scales are therefore prone to over-suppressing galaxies, as I-fronts can very quickly sweep through the gas and indefinitely prevent collapse until the potential well grows sufficiently large through merging.

The same is found by comparing numerical constraints on the characteristic mass scale, $M_{\mathrm{c}}$, with those of \cite{Okamoto:2008aa} whose numerical fits are commonly used to constrain $f_{\mathrm{b}}$ in SAMs of galaxy formation. In their models, homogeneous reionization is triggered instantaneously at $z_{\mathrm{reion}} = 9$, using the UV background model of \cite{Haardt:2001aa}. This assumption leads to a steep increase in $M_{\mathrm{c}}$, as haloes transition from unsuppressed to suppressed regime much faster than in our self-consistent models, where their suppression is dependent on their \emph{local} reionization history. The suppression continues to rise above our fiducial model due to a number of factors. First, contrary to our models, they neglect metal-line cooling, which denies the gas an efficient cooling mechanism at high densities. Secondly, due to their fixed number of smoothed particle hydrodynamics (SPH) particles, the smoothing length is only $\epsilon = 0.25\ \mathrm{kpc} h^{-1}$. This choice results in a maximum baryon overdensity of $\rho / \langle \rho \rangle \sim 10^{5}$ in their simulations, while we achieve $\rho / \langle \rho \rangle \sim 10^{8}$, which could reduce the ability of gas to self-shield efficiently. Finally, haloes which have undergone recent merger activity are removed from their results prior to fitting $M_{\mathrm{c}}$. As we have shown, strong tidal interactions leads to a large scatter, typically above the median, in $f_{\mathrm{b}}$. Removing these haloes from our sample leads to an increase in $M_{\mathrm{c}}$, partially easing the tension with our fiducial model. In contrast, our fiducial results are several orders of magnitude above the predictions of \cite{Hoeft:2006aa}, despite using the same SPH code {\sc gadget2} \citep{Springel:2005aa} and similar homogeneous UV background model of \cite{Haardt:1996aa} compared to that of \cite{Okamoto:2008aa}. In their model, reionization is not triggered until $z_{\mathrm{reion}} = 6$, hence their values initially lag behind those discussed above. Furthermore, they restrict their study to void galaxies, which are not representative of the wider galaxy population.

To improve on their models, we developed an ANN to predict the halo baryon fraction that is independent of redshift, mass, and crucially, reionization history. This model can account for the impact of mergers and tidal stripping, as we demonstrated by creating a mock catalogue of haloes with low tidal forces attributed to them. Both our simulations and ANN model predict a rise in $M_{\mathrm{c}}$ in such a scenario; however, there is mild tension between them due to the manner in which we randomly sample tidal forces to construct out training data. While the ANN reproduces much of the scatter in $f_{\mathrm{b}}$, marking a significant advancement over the current methods of SAMs that use only a mass- and redshift-dependent best-fit curve, it misses a small percentage of haloes with the highest baryon fractions. These objects require additional physics to constrain them within our model, which in turn would introduce additional degrees of freedom. One caveat of our model is that it requires the mass-weighted ionized fraction and gas temperature to be calculated for each halo, which may not be possible in some simpler SAMs. Therefore, we restrict the number of inputs (degrees of freedom) to four, to reduce the number of known quantities required to implement our model.

While our simulations are consistent with observational constraints of the volume-weighted hydrogen neutral fraction, photoionization rate, and Thompson scattering optical depth, there is the caveat that relatively small volumes exhibit a significant scatter in their EoR when examined as part of large-scale simulations \citep[see, for example][]{Ciardi:2003aa, Iliev:2006aa, Mellema:2006aa, Zahn:2007aa, McQuinn:2007aa, Iliev:2007ab, Alvarez:2007aa, Mesinger:2007aa, Geil:2008aa, Choudhury:2009aa}. While {\sc ramses-rt} employs a moment based method of RT, which scales independently of the number of sources and hence is not limited by volume, the code does not scale particularly well at our resolution of $\Delta x = 15.25\ \mathrm{pc}\ h^{-1}$. As such, the limiting factor in scaling our simulations up to larger volumes lies solely with access to large computational resources.

Due to computational constraints, we follow the methodology of \cite{Gnedin:2001ab} in reducing the speed of light to $f_{c} = 1/100$. This assumption allows us to relax the Courant-condition for the radiation fluid, thus reducing the total number of radiation sub-cycles and, hence, the total run-time (see also discussion in \citealt{Aubert:2008aa}). A framework for setting the reduced speed of light is set out in \cite{Rosdahl:2013aa}, following the study of relativistic I-fronts in \cite{Shapiro:2006aa}. A reduced speed of light increases the effective crossing of I-fronts, hence they will lag behind the full speed of light solution after a single light-crossing time (across the Str\"{o}mgren radius, in units of the recombination time). This effect can cause a retardation in the speed of I-fronts across cosmological volumes; therefore, we artificially boost the stellar escape fraction, $f^{*}_{\mathrm{esc}}$, in order to calibrate our reionization histories as shown in Fig.~\ref{fig:reion_hist}. While our minimum value is above unity, i.e. $f^{*}_{\mathrm{esc}} = 1.5$ for our RT1.5 simulation, we note that stellar synthesis uncertainties could give rise to much higher stellar luminosities than those predicted by \cite{Bruzual:2003aa} and used in our simulations (see the {\sc bpass} model, for example; \citealt{Eldridge:2008aa}). As we are primarily concerned with the properties of individual haloes in our simulations, and not the global statistics of reionization, the reduced speed of light approximation is valid as the effective crossing time of a halo is negligible when compared with the Universe.

\section{Conclusions}
\label{sec:conclusions}

We have presented the highest spatially resolved, fully self-consistent simulations of the halo baryon fraction during the EoR, achieving a formal resolution of $15.25\ \mathrm{pc} h^{-1}$ comoving using the {\sc ramses-rt} code \citep{Rosdahl:2013aa}. To disentangle the effects of tidal stripping, photoionization, and heating, we developed an Artificial Neural Network, which we trained using our fiducial simulation and then applied to three distinctly varying reionization histories. Our network is able to accurately reproduce the halo baryon fraction in each model, matching the measured value of the characteristic mass scale, $M_{\mathrm{c}}$ in each case (Fig.~\ref{fig:Mc_z_xHII_tidal_cutoff}).

Our findings are in disagreement with those of \cite{Okamoto:2008aa} and \cite{Hoeft:2006aa}, both of whom adopt the (instantaneous) homogeneous UV backgrounds of \cite{Haardt:2001aa} and \cite{Haardt:1996aa}, respectively. This discrepancy is partially due to stronger self-shielding in our models (as we reach much higher spatial resolutions and hence baryon overdensities) and an additional cooling mechanism available (i.e. efficient metal-line cooling at high densities). Our most extreme RT5 model, which is in tension observations of the neutral fraction and Thompson scattering optical depth constraints from Planck \citep{Planck-Collaboration:2016ab}, does match the \cite{Okamoto:2008aa} constraints by $z \sim 6$ and overshoots this value when haloes that have undergone tidal interactions are ignored.

Radiative feedback can suppress star formation in haloes, however not as strongly as previously suggested by \cite{Ocvirk:2016aa}. While the star formation rate in dwarf haloes (below the atomic cooling limit) is reduced, it is not quenched entirely. Furthermore, while the mean stellar abundance of haloes is reduced in our RT models, only our early reionization model (RT5) is inconsistent with the reference SNe-only case within $1 \sigma$. The degree of suppression in SFR is strongly correlated with the reduction in halo baryon fraction, as radiation primarily acts to prevent the collapse of star forming gas, rather than heat it past the temperature criterion for star formation. We find that modulation of star formation is primarily due to Jeans-mass filtering, preventing the collapse of baryons from the IGM into haloes and thus reducing their baryon fraction

The primary aim of this paper was to develop a model that can predict and capture the scatter in the baryon fraction of haloes in the suppressible regime, whose star formation rates are modulated during reionization. Existing models rely on fitting the average mass scale and slope that determines the transition from baryon poor to rich, providing a best-fit curve that misses much of the scatter in this quantity. We developed and trained an artificial neural network (ANN) to reproduce the halo baryon fraction based on four input parameters: the strongest tidal force exerted from a neighbour (averaged over a dynamical time), the (mass-weighted) hydrogen ionization fraction, the gas temperature (in units of the halo virial temperature), and the virial ratio (ratio of kinetic to potential energies). With these limited degrees of freedom, we demonstrated that our ANN could reproduce the characteristic mass scale, $M_{\mathrm{c}}$, independently of redshift, mass, and reionization history. Furthermore, it captures much of the scatter in the baryon fraction, leading to a more diverse galactic halo population.

Two of our input parameters (tidal force and virial ratio) can be computed purely from $N$-body simulations, and both are already provided by the {\sc rockstar} halo finder \citep{Behroozi:2013ab}. The remaining two quantities, the hydrogen ionization fraction and gas temperature can be approximated by semi-analytical models, see e.g. \cite{Srisawat:2016aa}. Therefore, our ANN can be trivially embedded within semi-analytical models of galaxy formation, leading to a more self-consistent calculation of the baryon fraction over current methods.

\section*{Acknowledgements}
The authors contributed in the following way to this paper. DS undertook the calibration and completion of all simulations; the development of all analysis codes, including the artificial neural network; the production of all figures and the first draft of the paper. II supervised the project and helped with the interpretation of results and feedback on the first draft. KD provided guidance throughout and significant feedback on the first draft.

DS would like to thank Joakim Rosdahl for valuable technical feedback, Scott J. Clay for advice and feedback on the first draft, Peter A. Thomas, Paul R. Shapiro and Beno\^{i}t Fournier for useful discussions and feedback.

This work used the DiRAC Data Centric system at Durham
University, operated by the Institute for Computational
Cosmology on behalf of the STFC DiRAC HPC Facility
(www.dirac.ac.uk). The DiRAC system is funded by
BIS National E-infrastructure capital grant ST/K00042X/1,
STFC capital grant ST/H008519/1, STFC DiRAC Operations
grant ST/K003267/1, and Durham University. DiRAC
is part of the National E-Infrastructure.
We acknowledge PRACE for awarding us computational time under project PRACE4LOFAR grant 2014102339 and ``Multi-scale simulations of Cosmic Reionization'' grants 2014102281, 2015122822, and 2016153528 to resource Curie based in France at CEA and to resource Marenostrum based in Spain at BSC. This work was supported by the Science and Technology Facilities Council [grant numbers ST/F002858/1 and ST/I000976/1] and the Southeast Physics Network (SEPNet). Some of the analysis was done on the Apollo cluster at The University of Sussex.

We made extensive use of the {\sc pymses} \citep{Guillet:2013aa} and {\sc pynbody} \citep{Pontzen:2013aa} python modules when developing our own analysis pipeline.




\bibliographystyle{mnras}
\bibliography{bib}

\begin{thebibliography}{}
\makeatletter
\relax
\def\mn@urlcharsother{\let\do\@makeother \do\$\do\&\do\#\do\^\do\_\do\%\do\~}
\def\mn@doi{\begingroup\mn@urlcharsother \@ifnextchar [ {\mn@doi@}
  {\mn@doi@[]}}
\def\mn@doi@[#1]#2{\def\@tempa{#1}\ifx\@tempa\@empty \href
  {http://dx.doi.org/#2} {doi:#2}\else \href {http://dx.doi.org/#2} {#1}\fi
  \endgroup}
\def\mn@eprint#1#2{\mn@eprint@#1:#2::\@nil}
\def\mn@eprint@arXiv#1{\href {http://arxiv.org/abs/#1} {{\tt arXiv:#1}}}
\def\mn@eprint@dblp#1{\href {http://dblp.uni-trier.de/rec/bibtex/#1.xml}
  {dblp:#1}}
\def\mn@eprint@#1:#2:#3:#4\@nil{\def\@tempa {#1}\def\@tempb {#2}\def\@tempc
  {#3}\ifx \@tempc \@empty \let \@tempc \@tempb \let \@tempb \@tempa \fi \ifx
  \@tempb \@empty \def\@tempb {arXiv}\fi \@ifundefined
  {mn@eprint@\@tempb}{\@tempb:\@tempc}{\expandafter \expandafter \csname
  mn@eprint@\@tempb\endcsname \expandafter{\@tempc}}}

\bibitem[\protect\citeauthoryear{{Abel}, {Anninos}, {Zhang}  \&
  {Norman}}{{Abel} et~al.}{1997}]{Abel:1997aa}
{Abel} T.,  {Anninos} P.,  {Zhang} Y.,   {Norman} M.~L.,  1997, \mn@doi [\na]
  {10.1016/S1384-1076(97)00010-9}, \href
  {http://adsabs.harvard.edu/abs/1997NewA....2..181A} {2, 181}

\bibitem[\protect\citeauthoryear{{Alvarez} \& {Abel}}{{Alvarez} \&
  {Abel}}{2007}]{Alvarez:2007aa}
{Alvarez} M.~A.,  {Abel} T.,  2007, \mn@doi [\mnras]
  {10.1111/j.1745-3933.2007.00342.x}, \href
  {http://adsabs.harvard.edu/abs/2007MNRAS.380L..30A} {380, L30}

\bibitem[\protect\citeauthoryear{{Arribas}, {Colina}, {Bellocchi}, {Maiolino}
  \& {Villar-Mart{\'{\i}}n}}{{Arribas} et~al.}{2014}]{Arribas:2014aa}
{Arribas} S.,  {Colina} L.,  {Bellocchi} E.,  {Maiolino} R.,
  {Villar-Mart{\'{\i}}n} M.,  2014, \mn@doi [\aap]
  {10.1051/0004-6361/201323324}, \href
  {http://adsabs.harvard.edu/abs/2014A%26A...568A..14A} {568, A14}

\bibitem[\protect\citeauthoryear{{Aubert} \& {Teyssier}}{{Aubert} \&
  {Teyssier}}{2008}]{Aubert:2008aa}
{Aubert} D.,  {Teyssier} R.,  2008, \mn@doi [\mnras]
  {10.1111/j.1365-2966.2008.13223.x}, \href
  {http://adsabs.harvard.edu/abs/2008MNRAS.387..295A} {387, 295}

\bibitem[\protect\citeauthoryear{{Aubert} \& {Teyssier}}{{Aubert} \&
  {Teyssier}}{2010}]{Aubert:2010aa}
{Aubert} D.,  {Teyssier} R.,  2010, \mn@doi [\apj]
  {10.1088/0004-637X/724/1/244}, \href
  {http://adsabs.harvard.edu/abs/2010ApJ...724..244A} {724, 244}

\bibitem[\protect\citeauthoryear{{Aubert}, {Deparis}  \& {Ocvirk}}{{Aubert}
  et~al.}{2015}]{Aubert:2015aa}
{Aubert} D.,  {Deparis} N.,   {Ocvirk} P.,  2015, \mn@doi [\mnras]
  {10.1093/mnras/stv1896}, \href
  {http://adsabs.harvard.edu/abs/2015MNRAS.454.1012A} {454, 1012}

\bibitem[\protect\citeauthoryear{{Babul} \& {Rees}}{{Babul} \&
  {Rees}}{1992}]{Babul:1992aa}
{Babul} A.,  {Rees} M.~J.,  1992, \mn@doi [\mnras] {10.1093/mnras/255.2.346},
  \href {http://adsabs.harvard.edu/abs/1992MNRAS.255..346B} {255, 346}

\bibitem[\protect\citeauthoryear{{Barkana} \& {Loeb}}{{Barkana} \&
  {Loeb}}{2000}]{Barkana:2000aa}
{Barkana} R.,  {Loeb} A.,  2000, \mn@doi [\apj] {10.1086/309229}, \href
  {http://adsabs.harvard.edu/abs/2000ApJ...539...20B} {539, 20}

\bibitem[\protect\citeauthoryear{{Behroozi}, {Wechsler}  \& {Wu}}{{Behroozi}
  et~al.}{2013a}]{Behroozi:2013ab}
{Behroozi} P.~S.,  {Wechsler} R.~H.,   {Wu} H.-Y.,  2013a, \mn@doi [\apj]
  {10.1088/0004-637X/762/2/109}, \href
  {http://adsabs.harvard.edu/abs/2013ApJ...762..109B} {762, 109}

\bibitem[\protect\citeauthoryear{{Behroozi}, {Wechsler}, {Wu}, {Busha},
  {Klypin}  \& {Primack}}{{Behroozi} et~al.}{2013b}]{Behroozi:2013ac}
{Behroozi} P.~S.,  {Wechsler} R.~H.,  {Wu} H.-Y.,  {Busha} M.~T.,  {Klypin}
  A.~A.,   {Primack} J.~R.,  2013b, \mn@doi [\apj]
  {10.1088/0004-637X/763/1/18}, \href
  {http://adsabs.harvard.edu/abs/2013ApJ...763...18B} {763, 18}

\bibitem[\protect\citeauthoryear{{Behroozi}, {Wechsler}  \&
  {Conroy}}{{Behroozi} et~al.}{2013c}]{Behroozi:2013aa}
{Behroozi} P.~S.,  {Wechsler} R.~H.,   {Conroy} C.,  2013c, \mn@doi [\apj]
  {10.1088/0004-637X/770/1/57}, \href
  {http://adsabs.harvard.edu/abs/2013ApJ...770...57B} {770, 57}

\bibitem[\protect\citeauthoryear{{Belokurov} et~al.,}{{Belokurov}
  et~al.}{2006}]{Belokurov:2006aa}
{Belokurov} V.,  et~al., 2006, \mn@doi [\apjl] {10.1086/504797}, \href
  {http://adsabs.harvard.edu/abs/2006ApJ...642L.137B} {642, L137}

\bibitem[\protect\citeauthoryear{{Benson}, {Lacey}, {Baugh}, {Cole}  \&
  {Frenk}}{{Benson} et~al.}{2002a}]{Benson:2002aa}
{Benson} A.~J.,  {Lacey} C.~G.,  {Baugh} C.~M.,  {Cole} S.,   {Frenk} C.~S.,
  2002a, \mn@doi [\mnras] {10.1046/j.1365-8711.2002.05387.x}, \href
  {http://adsabs.harvard.edu/abs/2002MNRAS.333..156B} {333, 156}

\bibitem[\protect\citeauthoryear{{Benson}, {Frenk}, {Lacey}, {Baugh}  \&
  {Cole}}{{Benson} et~al.}{2002b}]{Benson:2002ab}
{Benson} A.~J.,  {Frenk} C.~S.,  {Lacey} C.~G.,  {Baugh} C.~M.,   {Cole} S.,
  2002b, \mn@doi [\mnras] {10.1046/j.1365-8711.2002.05388.x}, \href
  {http://adsabs.harvard.edu/abs/2002MNRAS.333..177B} {333, 177}

\bibitem[\protect\citeauthoryear{{Bruzual} \& {Charlot}}{{Bruzual} \&
  {Charlot}}{2003}]{Bruzual:2003aa}
{Bruzual} G.,  {Charlot} S.,  2003, \mn@doi [\mnras]
  {10.1046/j.1365-8711.2003.06897.x}, \href
  {http://adsabs.harvard.edu/abs/2003MNRAS.344.1000B} {344, 1000}

\bibitem[\protect\citeauthoryear{{Bryan} \& {Norman}}{{Bryan} \&
  {Norman}}{1998}]{Bryan:1998aa}
{Bryan} G.~L.,  {Norman} M.~L.,  1998, \mn@doi [\apj] {10.1086/305262}, \href
  {http://adsabs.harvard.edu/abs/1998ApJ...495...80B} {495, 80}

\bibitem[\protect\citeauthoryear{{Calverley}, {Becker}, {Haehnelt}  \&
  {Bolton}}{{Calverley} et~al.}{2011}]{Calverley:2011aa}
{Calverley} A.~P.,  {Becker} G.~D.,  {Haehnelt} M.~G.,   {Bolton} J.~S.,  2011,
  \mn@doi [\mnras] {10.1111/j.1365-2966.2010.18072.x}, \href
  {http://adsabs.harvard.edu/abs/2011MNRAS.412.2543C} {412, 2543}

\bibitem[\protect\citeauthoryear{{Chabrier}}{{Chabrier}}{2003}]{Chabrier:2003aa}
{Chabrier} G.,  2003, \mn@doi [\pasp] {10.1086/376392}, \href
  {http://adsabs.harvard.edu/abs/2003PASP..115..763C} {115, 763}

\bibitem[\protect\citeauthoryear{{Chisholm}, {Tremonti}, {Leitherer}, {Chen},
  {Wofford}  \& {Lundgren}}{{Chisholm} et~al.}{2015}]{Chisholm:2015aa}
{Chisholm} J.,  {Tremonti} C.~A.,  {Leitherer} C.,  {Chen} Y.,  {Wofford} A.,
  {Lundgren} B.,  2015, \mn@doi [\apj] {10.1088/0004-637X/811/2/149}, \href
  {http://adsabs.harvard.edu/abs/2015ApJ...811..149C} {811, 149}

\bibitem[\protect\citeauthoryear{{Chornock}, {Berger}, {Fox}, {Lunnan},
  {Drout}, {Fong}, {Laskar}  \& {Roth}}{{Chornock}
  et~al.}{2013}]{Chornock:2013aa}
{Chornock} R.,  {Berger} E.,  {Fox} D.~B.,  {Lunnan} R.,  {Drout} M.~R.,
  {Fong} W.-f.,  {Laskar} T.,   {Roth} K.~C.,  2013, \mn@doi [\apj]
  {10.1088/0004-637X/774/1/26}, \href
  {http://adsabs.harvard.edu/abs/2013ApJ...774...26C} {774, 26}

\bibitem[\protect\citeauthoryear{{Choudhury}, {Haehnelt}  \&
  {Regan}}{{Choudhury} et~al.}{2009}]{Choudhury:2009aa}
{Choudhury} T.~R.,  {Haehnelt} M.~G.,   {Regan} J.,  2009, \mn@doi [\mnras]
  {10.1111/j.1365-2966.2008.14383.x}, \href
  {http://adsabs.harvard.edu/abs/2009MNRAS.394..960C} {394, 960}

\bibitem[\protect\citeauthoryear{{Ciardi}, {Stoehr}  \& {White}}{{Ciardi}
  et~al.}{2003a}]{Ciardi:2003ab}
{Ciardi} B.,  {Stoehr} F.,   {White} S.~D.~M.,  2003a, \mn@doi [\mnras]
  {10.1046/j.1365-8711.2003.06797.x}, \href
  {http://adsabs.harvard.edu/abs/2003MNRAS.343.1101C} {343, 1101}

\bibitem[\protect\citeauthoryear{{Ciardi}, {Ferrara}  \& {White}}{{Ciardi}
  et~al.}{2003b}]{Ciardi:2003aa}
{Ciardi} B.,  {Ferrara} A.,   {White} S.~D.~M.,  2003b, \mn@doi [\mnras]
  {10.1046/j.1365-8711.2003.06976.x}, \href
  {http://adsabs.harvard.edu/abs/2003MNRAS.344L...7C} {344, L7}

\bibitem[\protect\citeauthoryear{{Clark}, {Glover}, {Klessen}  \&
  {Bromm}}{{Clark} et~al.}{2011}]{Clark:2011aa}
{Clark} P.~C.,  {Glover} S.~C.~O.,  {Klessen} R.~S.,   {Bromm} V.,  2011,
  \mn@doi [\apj] {10.1088/0004-637X/727/2/110}, \href
  {http://adsabs.harvard.edu/abs/2011ApJ...727..110C} {727, 110}

\bibitem[\protect\citeauthoryear{{Couchman} \& {Rees}}{{Couchman} \&
  {Rees}}{1986}]{Couchman:1986aa}
{Couchman} H.~M.~P.,  {Rees} M.~J.,  1986, \mn@doi [\mnras]
  {10.1093/mnras/221.1.53}, \href
  {http://adsabs.harvard.edu/abs/1986MNRAS.221...53C} {221, 53}

\bibitem[\protect\citeauthoryear{Cybenko}{Cybenko}{1989}]{Cybenko1989}
Cybenko G.,  1989, Mathematics of Control, Signals, and Systems (MCSS), 2, 303

\bibitem[\protect\citeauthoryear{{Davis}, {D'Aloisio}  \& {Natarajan}}{{Davis}
  et~al.}{2011}]{Davis:2011aa}
{Davis} A.~J.,  {D'Aloisio} A.,   {Natarajan} P.,  2011, \mn@doi [\mnras]
  {10.1111/j.1365-2966.2011.19026.x}, \href
  {http://adsabs.harvard.edu/abs/2011MNRAS.416..242D} {416, 242}

\bibitem[\protect\citeauthoryear{{Dixon}, {Iliev}, {Mellema}, {Ahn}  \&
  {Shapiro}}{{Dixon} et~al.}{2016}]{Dixon:2016aa}
{Dixon} K.~L.,  {Iliev} I.~T.,  {Mellema} G.,  {Ahn} K.,   {Shapiro} P.~R.,
  2016, \mn@doi [\mnras] {10.1093/mnras/stv2887}, \href
  {http://adsabs.harvard.edu/abs/2016MNRAS.456.3011D} {456, 3011}

\bibitem[\protect\citeauthoryear{{Doroshkevich}, {Zel'dovich}  \&
  {Novikov}}{{Doroshkevich} et~al.}{1967}]{Doroshkevich:1967aa}
{Doroshkevich} A.~G.,  {Zel'dovich} Y.~B.,   {Novikov} I.~D.,  1967, \sovast,
  \href {http://adsabs.harvard.edu/abs/1967SvA....11..233D} {11, 233}

\bibitem[\protect\citeauthoryear{{Dubois} \& {Teyssier}}{{Dubois} \&
  {Teyssier}}{2008}]{Dubois:2008aa}
{Dubois} Y.,  {Teyssier} R.,  2008, \mn@doi [\aap]
  {10.1051/0004-6361:20078326}, \href
  {http://adsabs.harvard.edu/abs/2008A%26A...477...79D} {477, 79}

\bibitem[\protect\citeauthoryear{{Efstathiou}}{{Efstathiou}}{1992}]{Efstathiou:1992aa}
{Efstathiou} G.,  1992, \mn@doi [\mnras] {10.1093/mnras/256.1.43P}, \href
  {http://adsabs.harvard.edu/abs/1992MNRAS.256P..43E} {256, 43P}

\bibitem[\protect\citeauthoryear{{Eldridge}, {Izzard}  \& {Tout}}{{Eldridge}
  et~al.}{2008}]{Eldridge:2008aa}
{Eldridge} J.~J.,  {Izzard} R.~G.,   {Tout} C.~A.,  2008, \mn@doi [\mnras]
  {10.1111/j.1365-2966.2007.12738.x}, \href
  {http://adsabs.harvard.edu/abs/2008MNRAS.384.1109E} {384, 1109}

\bibitem[\protect\citeauthoryear{{Fan} et~al.,}{{Fan}
  et~al.}{2006}]{Fan:2006aa}
{Fan} X.,  et~al., 2006, \mn@doi [\aj] {10.1086/504836}, \href
  {http://adsabs.harvard.edu/abs/2006AJ....132..117F} {132, 117}

\bibitem[\protect\citeauthoryear{{Finlator}, {{\"O}zel}, {Dav{\'e}}  \&
  {Oppenheimer}}{{Finlator} et~al.}{2009}]{Finlator:2009aa}
{Finlator} K.,  {{\"O}zel} F.,  {Dav{\'e}} R.,   {Oppenheimer} B.~D.,  2009,
  \mn@doi [\mnras] {10.1111/j.1365-2966.2009.15521.x}, \href
  {http://adsabs.harvard.edu/abs/2009MNRAS.400.1049F} {400, 1049}

\bibitem[\protect\citeauthoryear{{Geil} \& {Wyithe}}{{Geil} \&
  {Wyithe}}{2008}]{Geil:2008aa}
{Geil} P.~M.,  {Wyithe} J.~S.~B.,  2008, \mn@doi [\mnras]
  {10.1111/j.1365-2966.2008.13159.x}, \href
  {http://adsabs.harvard.edu/abs/2008MNRAS.386.1683G} {386, 1683}

\bibitem[\protect\citeauthoryear{{Gnedin}}{{Gnedin}}{2000a}]{Gnedin:2000ab}
{Gnedin} N.~Y.,  2000a, \mn@doi [\apj] {10.1086/308876}, \href
  {http://adsabs.harvard.edu/abs/2000ApJ...535..530G} {535, 530}

\bibitem[\protect\citeauthoryear{{Gnedin}}{{Gnedin}}{2000b}]{Gnedin:2000aa}
{Gnedin} N.~Y.,  2000b, \mn@doi [\apj] {10.1086/317042}, \href
  {http://adsabs.harvard.edu/abs/2000ApJ...542..535G} {542, 535}

\bibitem[\protect\citeauthoryear{{Gnedin} \& {Abel}}{{Gnedin} \&
  {Abel}}{2001}]{Gnedin:2001ab}
{Gnedin} N.~Y.,  {Abel} T.,  2001, \mn@doi [\na]
  {10.1016/S1384-1076(01)00068-9}, \href
  {http://adsabs.harvard.edu/abs/2001NewA....6..437G} {6, 437}

\bibitem[\protect\citeauthoryear{{Gnedin} \& {Ostriker}}{{Gnedin} \&
  {Ostriker}}{1997}]{Gnedin:1997aa}
{Gnedin} N.~Y.,  {Ostriker} J.~P.,  1997, \mn@doi [\apj] {10.1086/304548},
  \href {http://adsabs.harvard.edu/abs/1997ApJ...486..581G} {486, 581}

\bibitem[\protect\citeauthoryear{{Greif}, {White}, {Klessen}  \&
  {Springel}}{{Greif} et~al.}{2011}]{Greif:2011aa}
{Greif} T.~H.,  {White} S.~D.~M.,  {Klessen} R.~S.,   {Springel} V.,  2011,
  \mn@doi [\apj] {10.1088/0004-637X/736/2/147}, \href
  {http://adsabs.harvard.edu/abs/2011ApJ...736..147G} {736, 147}

\bibitem[\protect\citeauthoryear{{Guillet}, {Chapon}  \& {Labadens}}{{Guillet}
  et~al.}{2013}]{Guillet:2013aa}
{Guillet} T.,  {Chapon} D.,   {Labadens} M.,  2013, {PyMSES: Python modules for
  RAMSES}, Astrophysics Source Code Library (\mn@eprint {ascl} {1310.002})

\bibitem[\protect\citeauthoryear{{Haardt} \& {Madau}}{{Haardt} \&
  {Madau}}{1996}]{Haardt:1996aa}
{Haardt} F.,  {Madau} P.,  1996, \mn@doi [\apj] {10.1086/177035}, \href
  {http://adsabs.harvard.edu/abs/1996ApJ...461...20H} {461, 20}

\bibitem[\protect\citeauthoryear{{Haardt} \& {Madau}}{{Haardt} \&
  {Madau}}{2001}]{Haardt:2001aa}
{Haardt} F.,  {Madau} P.,  2001, in {Neumann} D.~M.,  {Tran} J.~T.~V.,  eds,
  Clusters of Galaxies and the High Redshift Universe Observed in X-rays.
  (\mn@eprint {} {astro-ph/0106018})

\bibitem[\protect\citeauthoryear{{Hahn} \& {Abel}}{{Hahn} \&
  {Abel}}{2013}]{Hahn:2013aa}
{Hahn} O.,  {Abel} T.,  2013, {MUSIC: MUlti-Scale Initial Conditions},
  Astrophysics Source Code Library (\mn@eprint {ascl} {1311.011})

\bibitem[\protect\citeauthoryear{{Henriques}, {White}, {Thomas}, {Angulo},
  {Guo}, {Lemson}, {Springel}  \& {Overzier}}{{Henriques}
  et~al.}{2015}]{Henriques:2015aa}
{Henriques} B.~M.~B.,  {White} S.~D.~M.,  {Thomas} P.~A.,  {Angulo} R.,  {Guo}
  Q.,  {Lemson} G.,  {Springel} V.,   {Overzier} R.,  2015, \mn@doi [\mnras]
  {10.1093/mnras/stv705}, \href
  {http://adsabs.harvard.edu/abs/2015MNRAS.451.2663H} {451, 2663}

\bibitem[\protect\citeauthoryear{{Hoeft}, {Yepes}, {Gottl{\"o}ber}  \&
  {Springel}}{{Hoeft} et~al.}{2006}]{Hoeft:2006aa}
{Hoeft} M.,  {Yepes} G.,  {Gottl{\"o}ber} S.,   {Springel} V.,  2006, \mn@doi
  [\mnras] {10.1111/j.1365-2966.2006.10678.x}, \href
  {http://adsabs.harvard.edu/abs/2006MNRAS.371..401H} {371, 401}

\bibitem[\protect\citeauthoryear{Hornik, Stinchcombe  \& White}{Hornik
  et~al.}{1989}]{Hornik1989}
Hornik K.,  Stinchcombe M.,   White H.,  1989, Neural Networks, 2, 359

\bibitem[\protect\citeauthoryear{{Iliev}, {Shapiro}  \& {Raga}}{{Iliev}
  et~al.}{2005a}]{Iliev:2005ab}
{Iliev} I.~T.,  {Shapiro} P.~R.,   {Raga} A.~C.,  2005a, \mn@doi [\mnras]
  {10.1111/j.1365-2966.2005.09155.x}, \href
  {http://adsabs.harvard.edu/abs/2005MNRAS.361..405I} {361, 405}

\bibitem[\protect\citeauthoryear{{Iliev}, {Scannapieco}  \& {Shapiro}}{{Iliev}
  et~al.}{2005b}]{Iliev:2005aa}
{Iliev} I.~T.,  {Scannapieco} E.,   {Shapiro} P.~R.,  2005b, \mn@doi [\apj]
  {10.1086/429083}, \href {http://adsabs.harvard.edu/abs/2005ApJ...624..491I}
  {624, 491}

\bibitem[\protect\citeauthoryear{{Iliev}, {Mellema}, {Pen}, {Merz}, {Shapiro}
  \& {Alvarez}}{{Iliev} et~al.}{2006}]{Iliev:2006aa}
{Iliev} I.~T.,  {Mellema} G.,  {Pen} U.-L.,  {Merz} H.,  {Shapiro} P.~R.,
  {Alvarez} M.~A.,  2006, \mn@doi [\mnras] {10.1111/j.1365-2966.2006.10502.x},
  \href {http://adsabs.harvard.edu/abs/2006MNRAS.369.1625I} {369, 1625}

\bibitem[\protect\citeauthoryear{{Iliev}, {Mellema}, {Shapiro}  \&
  {Pen}}{{Iliev} et~al.}{2007}]{Iliev:2007ab}
{Iliev} I.~T.,  {Mellema} G.,  {Shapiro} P.~R.,   {Pen} U.-L.,  2007, \mn@doi
  [\mnras] {10.1111/j.1365-2966.2007.11482.x}, \href
  {http://adsabs.harvard.edu/abs/2007MNRAS.376..534I} {376, 534}

\bibitem[\protect\citeauthoryear{{Iliev}, {Mellema}, {Ahn}, {Shapiro}, {Mao}
  \& {Pen}}{{Iliev} et~al.}{2014}]{Iliev:2014aa}
{Iliev} I.~T.,  {Mellema} G.,  {Ahn} K.,  {Shapiro} P.~R.,  {Mao} Y.,   {Pen}
  U.-L.,  2014, \mn@doi [\mnras] {10.1093/mnras/stt2497}, \href
  {http://adsabs.harvard.edu/abs/2014MNRAS.439..725I} {439, 725}

\bibitem[\protect\citeauthoryear{{Khokhlov}}{{Khokhlov}}{1998}]{Khokhlov:1998aa}
{Khokhlov} A.,  1998, \mn@doi [Journal of Computational Physics]
  {10.1006/jcph.1998.9998}, \href
  {http://adsabs.harvard.edu/abs/1998JCoPh.143..519K} {143, 519}

\bibitem[\protect\citeauthoryear{{Kimm} \& {Cen}}{{Kimm} \&
  {Cen}}{2014}]{Kimm:2014aa}
{Kimm} T.,  {Cen} R.,  2014, \mn@doi [\apj] {10.1088/0004-637X/788/2/121},
  \href {http://adsabs.harvard.edu/abs/2014ApJ...788..121K} {788, 121}

\bibitem[\protect\citeauthoryear{{Komatsu} et~al.,}{{Komatsu}
  et~al.}{2011}]{Komatsu:2011aa}
{Komatsu} E.,  et~al., 2011, \mn@doi [\apjs] {10.1088/0067-0049/192/2/18},
  \href {http://adsabs.harvard.edu/abs/2011ApJS..192...18K} {192, 18}

\bibitem[\protect\citeauthoryear{{Lacey} et~al.,}{{Lacey}
  et~al.}{2016}]{Lacey:2016aa}
{Lacey} C.~G.,  et~al., 2016, \mn@doi [\mnras] {10.1093/mnras/stw1888}, \href
  {http://adsabs.harvard.edu/abs/2016MNRAS.462.3854L} {462, 3854}

\bibitem[\protect\citeauthoryear{{Martin}}{{Martin}}{2005}]{Martin:2005aa}
{Martin} C.~L.,  2005, \mn@doi [\apj] {10.1086/427277}, \href
  {http://adsabs.harvard.edu/abs/2005ApJ...621..227M} {621, 227}

\bibitem[\protect\citeauthoryear{{Martin}, {Shapley}, {Coil}, {Kornei},
  {Bundy}, {Weiner}, {Noeske}  \& {Schiminovich}}{{Martin}
  et~al.}{2012}]{Martin:2012aa}
{Martin} C.~L.,  {Shapley} A.~E.,  {Coil} A.~L.,  {Kornei} K.~A.,  {Bundy} K.,
  {Weiner} B.~J.,  {Noeske} K.~G.,   {Schiminovich} D.,  2012, \mn@doi [\apj]
  {10.1088/0004-637X/760/2/127}, \href
  {http://adsabs.harvard.edu/abs/2012ApJ...760..127M} {760, 127}

\bibitem[\protect\citeauthoryear{{McGreer}, {Mesinger}  \& {Fan}}{{McGreer}
  et~al.}{2011}]{McGreer:2011aa}
{McGreer} I.~D.,  {Mesinger} A.,   {Fan} X.,  2011, \mn@doi [\mnras]
  {10.1111/j.1365-2966.2011.18935.x}, \href
  {http://adsabs.harvard.edu/abs/2011MNRAS.415.3237M} {415, 3237}

\bibitem[\protect\citeauthoryear{{McGreer}, {Mesinger}  \&
  {D'Odorico}}{{McGreer} et~al.}{2015}]{McGreer:2015aa}
{McGreer} I.~D.,  {Mesinger} A.,   {D'Odorico} V.,  2015, \mn@doi [\mnras]
  {10.1093/mnras/stu2449}, \href
  {http://adsabs.harvard.edu/abs/2015MNRAS.447..499M} {447, 499}

\bibitem[\protect\citeauthoryear{{McKee} \& {Ostriker}}{{McKee} \&
  {Ostriker}}{1977}]{McKee:1977aa}
{McKee} C.~F.,  {Ostriker} J.~P.,  1977, \mn@doi [\apj] {10.1086/155667}, \href
  {http://adsabs.harvard.edu/abs/1977ApJ...218..148M} {218, 148}

\bibitem[\protect\citeauthoryear{{McQuinn}, {Lidz}, {Zahn}, {Dutta},
  {Hernquist}  \& {Zaldarriaga}}{{McQuinn} et~al.}{2007}]{McQuinn:2007aa}
{McQuinn} M.,  {Lidz} A.,  {Zahn} O.,  {Dutta} S.,  {Hernquist} L.,
  {Zaldarriaga} M.,  2007, \mn@doi [\mnras] {10.1111/j.1365-2966.2007.11489.x},
  \href {http://adsabs.harvard.edu/abs/2007MNRAS.377.1043M} {377, 1043}

\bibitem[\protect\citeauthoryear{{McQuinn}, {Lidz}, {Zaldarriaga}, {Hernquist}
  \& {Dutta}}{{McQuinn} et~al.}{2008}]{McQuinn:2008aa}
{McQuinn} M.,  {Lidz} A.,  {Zaldarriaga} M.,  {Hernquist} L.,   {Dutta} S.,
  2008, \mn@doi [\mnras] {10.1111/j.1365-2966.2008.13271.x}, \href
  {http://adsabs.harvard.edu/abs/2008MNRAS.388.1101M} {388, 1101}

\bibitem[\protect\citeauthoryear{{Mellema}, {Iliev}, {Alvarez}  \&
  {Shapiro}}{{Mellema} et~al.}{2006a}]{Mellema:2006ab}
{Mellema} G.,  {Iliev} I.~T.,  {Alvarez} M.~A.,   {Shapiro} P.~R.,  2006a,
  \mn@doi [\na] {10.1016/j.newast.2005.09.004}, \href
  {http://adsabs.harvard.edu/abs/2006NewA...11..374M} {11, 374}

\bibitem[\protect\citeauthoryear{{Mellema}, {Iliev}, {Pen}  \&
  {Shapiro}}{{Mellema} et~al.}{2006b}]{Mellema:2006aa}
{Mellema} G.,  {Iliev} I.~T.,  {Pen} U.-L.,   {Shapiro} P.~R.,  2006b, \mn@doi
  [\mnras] {10.1111/j.1365-2966.2006.10919.x}, \href
  {http://adsabs.harvard.edu/abs/2006MNRAS.372..679M} {372, 679}

\bibitem[\protect\citeauthoryear{{Mesinger} \& {Furlanetto}}{{Mesinger} \&
  {Furlanetto}}{2007}]{Mesinger:2007aa}
{Mesinger} A.,  {Furlanetto} S.,  2007, \mn@doi [\apj] {10.1086/521806}, \href
  {http://adsabs.harvard.edu/abs/2007ApJ...669..663M} {669, 663}

\bibitem[\protect\citeauthoryear{{Moore}, {Ghigna}, {Governato}, {Lake},
  {Quinn}, {Stadel}  \& {Tozzi}}{{Moore} et~al.}{1999}]{Moore:1999aa}
{Moore} B.,  {Ghigna} S.,  {Governato} F.,  {Lake} G.,  {Quinn} T.,  {Stadel}
  J.,   {Tozzi} P.,  1999, \mn@doi [\apjl] {10.1086/312287}, \href
  {http://adsabs.harvard.edu/abs/1999ApJ...524L..19M} {524, L19}

\bibitem[\protect\citeauthoryear{{Moster}, {Naab}  \& {White}}{{Moster}
  et~al.}{2013}]{Moster:2013aa}
{Moster} B.~P.,  {Naab} T.,   {White} S.~D.~M.,  2013, \mn@doi [\mnras]
  {10.1093/mnras/sts261}, \href
  {http://adsabs.harvard.edu/abs/2013MNRAS.428.3121M} {428, 3121}

\bibitem[\protect\citeauthoryear{{Nagashima}, {Gouda}  \&
  {Sugiura}}{{Nagashima} et~al.}{1999}]{Nagashima:1999aa}
{Nagashima} M.,  {Gouda} N.,   {Sugiura} N.,  1999, \mn@doi [\mnras]
  {10.1046/j.1365-8711.1999.02438.x}, \href
  {http://adsabs.harvard.edu/abs/1999MNRAS.305..449N} {305, 449}

\bibitem[\protect\citeauthoryear{{Nakamoto}, {Umemura}  \& {Susa}}{{Nakamoto}
  et~al.}{2001}]{Nakamoto:2001aa}
{Nakamoto} T.,  {Umemura} M.,   {Susa} H.,  2001, \mn@doi [\mnras]
  {10.1046/j.1365-8711.2001.04008.x}, \href
  {http://adsabs.harvard.edu/abs/2001MNRAS.321..593N} {321, 593}

\bibitem[\protect\citeauthoryear{{Newman} et~al.,}{{Newman}
  et~al.}{2012}]{Newman:2012aa}
{Newman} S.~F.,  et~al., 2012, \mn@doi [\apj] {10.1088/0004-637X/761/1/43},
  \href {http://adsabs.harvard.edu/abs/2012ApJ...761...43N} {761, 43}

\bibitem[\protect\citeauthoryear{{Noh} \& {McQuinn}}{{Noh} \&
  {McQuinn}}{2014}]{Noh:2014aa}
{Noh} Y.,  {McQuinn} M.,  2014, \mn@doi [\mnras] {10.1093/mnras/stu1412}, \href
  {http://adsabs.harvard.edu/abs/2014MNRAS.444..503N} {444, 503}

\bibitem[\protect\citeauthoryear{{Ocvirk} et~al.,}{{Ocvirk}
  et~al.}{2014}]{Ocvirk:2014aa}
{Ocvirk} P.,  et~al., 2014, \mn@doi [\apj] {10.1088/0004-637X/794/1/20}, \href
  {http://adsabs.harvard.edu/abs/2014ApJ...794...20O} {794, 20}

\bibitem[\protect\citeauthoryear{{Ocvirk} et~al.,}{{Ocvirk}
  et~al.}{2016}]{Ocvirk:2016aa}
{Ocvirk} P.,  et~al., 2016, \mn@doi [\mnras] {10.1093/mnras/stw2036}, \href
  {http://adsabs.harvard.edu/abs/2016MNRAS.463.1462O} {463, 1462}

\bibitem[\protect\citeauthoryear{{Okamoto}, {Gao}  \& {Theuns}}{{Okamoto}
  et~al.}{2008}]{Okamoto:2008aa}
{Okamoto} T.,  {Gao} L.,   {Theuns} T.,  2008, \mn@doi [\mnras]
  {10.1111/j.1365-2966.2008.13830.x}, \href
  {http://adsabs.harvard.edu/abs/2008MNRAS.390..920O} {390, 920}

\bibitem[\protect\citeauthoryear{{Ota} et~al.,}{{Ota}
  et~al.}{2008}]{Ota:2008aa}
{Ota} K.,  et~al., 2008, \mn@doi [\apj] {10.1086/529006}, \href
  {http://adsabs.harvard.edu/abs/2008ApJ...677...12O} {677, 12}

\bibitem[\protect\citeauthoryear{{Ouchi} et~al.,}{{Ouchi}
  et~al.}{2010}]{Ouchi:2010aa}
{Ouchi} M.,  et~al., 2010, \mn@doi [\apj] {10.1088/0004-637X/723/1/869}, \href
  {http://adsabs.harvard.edu/abs/2010ApJ...723..869O} {723, 869}

\bibitem[\protect\citeauthoryear{{Pawlik} \& {Schaye}}{{Pawlik} \&
  {Schaye}}{2009}]{Pawlik:2009aa}
{Pawlik} A.~H.,  {Schaye} J.,  2009, \mn@doi [\mnras]
  {10.1111/j.1745-3933.2009.00659.x}, \href
  {http://adsabs.harvard.edu/abs/2009MNRAS.396L..46P} {396, L46}

\bibitem[\protect\citeauthoryear{{Pawlik}, {Schaye}  \& {Dalla
  Vecchia}}{{Pawlik} et~al.}{2015}]{Pawlik:2015aa}
{Pawlik} A.~H.,  {Schaye} J.,   {Dalla Vecchia} C.,  2015, \mn@doi [\mnras]
  {10.1093/mnras/stv976}, \href
  {http://adsabs.harvard.edu/abs/2015MNRAS.451.1586P} {451, 1586}

\bibitem[\protect\citeauthoryear{{Planck Collaboration} et~al.,}{{Planck
  Collaboration} et~al.}{2016a}]{Planck-Collaboration:2016aa}
{Planck Collaboration} et~al., 2016a, \mn@doi [\aap]
  {10.1051/0004-6361/201525830}, \href
  {http://adsabs.harvard.edu/abs/2016A%26A...594A..13P} {594, A13}

\bibitem[\protect\citeauthoryear{{Planck Collaboration} et~al.,}{{Planck
  Collaboration} et~al.}{2016b}]{Planck-Collaboration:2016ab}
{Planck Collaboration} et~al., 2016b, \mn@doi [\aap]
  {10.1051/0004-6361/201628897}, \href
  {http://adsabs.harvard.edu/abs/2016A%26A...596A.108P} {596, A108}

\bibitem[\protect\citeauthoryear{{Pontzen}, {Ro{\v s}kar}, {Stinson}  \&
  {Woods}}{{Pontzen} et~al.}{2013}]{Pontzen:2013aa}
{Pontzen} A.,  {Ro{\v s}kar} R.,  {Stinson} G.,   {Woods} R.,  2013, {pynbody:
  N-Body/SPH analysis for python}, Astrophysics Source Code Library (\mn@eprint
  {ascl} {1305.002})

\bibitem[\protect\citeauthoryear{{Quinn}, {Katz}  \& {Efstathiou}}{{Quinn}
  et~al.}{1996}]{Quinn:1996aa}
{Quinn} T.,  {Katz} N.,   {Efstathiou} G.,  1996, \mn@doi [\mnras]
  {10.1093/mnras/278.4.L49}, \href
  {http://adsabs.harvard.edu/abs/1996MNRAS.278L..49Q} {278, L49}

\bibitem[\protect\citeauthoryear{{Rasera} \& {Teyssier}}{{Rasera} \&
  {Teyssier}}{2006}]{Rasera:2006aa}
{Rasera} Y.,  {Teyssier} R.,  2006, \mn@doi [\aap]
  {10.1051/0004-6361:20053116}, \href
  {http://adsabs.harvard.edu/abs/2006A%26A...445....1R} {445, 1}

\bibitem[\protect\citeauthoryear{{Razoumov} \& {Sommer-Larsen}}{{Razoumov} \&
  {Sommer-Larsen}}{2006}]{Razoumov:2006aa}
{Razoumov} A.~O.,  {Sommer-Larsen} J.,  2006, \mn@doi [\apjl] {10.1086/509636},
  \href {http://adsabs.harvard.edu/abs/2006ApJ...651L..89R} {651, L89}

\bibitem[\protect\citeauthoryear{{Rees}}{{Rees}}{1999}]{Rees:1999aa}
{Rees} M.~J.,  1999, in {Holt} S.,  {Smith} E.,  eds,  American Institute of
  Physics Conference Series Vol. 470, After the Dark Ages: When Galaxies were
  Young (the Universe at 2 $\lt$ Z $\lt$ 5). pp 13--23,
  \mn@doi{10.1063/1.58643}

\bibitem[\protect\citeauthoryear{{Robertson}, {Ellis}, {Furlanetto}  \&
  {Dunlop}}{{Robertson} et~al.}{2015}]{Robertson:2015aa}
{Robertson} B.~E.,  {Ellis} R.~S.,  {Furlanetto} S.~R.,   {Dunlop} J.~S.,
  2015, \mn@doi [\apjl] {10.1088/2041-8205/802/2/L19}, \href
  {http://adsabs.harvard.edu/abs/2015ApJ...802L..19R} {802, L19}

\bibitem[\protect\citeauthoryear{{Rosdahl} \& {Blaizot}}{{Rosdahl} \&
  {Blaizot}}{2012}]{Rosdahl:2012aa}
{Rosdahl} J.,  {Blaizot} J.,  2012, \mn@doi [\mnras]
  {10.1111/j.1365-2966.2012.20883.x}, \href
  {http://adsabs.harvard.edu/abs/2012MNRAS.423..344R} {423, 344}

\bibitem[\protect\citeauthoryear{{Rosdahl}, {Blaizot}, {Aubert}, {Stranex}  \&
  {Teyssier}}{{Rosdahl} et~al.}{2013}]{Rosdahl:2013aa}
{Rosdahl} J.,  {Blaizot} J.,  {Aubert} D.,  {Stranex} T.,   {Teyssier} R.,
  2013, \mn@doi [\mnras] {10.1093/mnras/stt1722}, \href
  {http://adsabs.harvard.edu/abs/2013MNRAS.436.2188R} {436, 2188}

\bibitem[\protect\citeauthoryear{{Rosdahl}, {Schaye}, {Teyssier}  \&
  {Agertz}}{{Rosdahl} et~al.}{2015}]{Rosdahl:2015aa}
{Rosdahl} J.,  {Schaye} J.,  {Teyssier} R.,   {Agertz} O.,  2015, \mn@doi
  [\mnras] {10.1093/mnras/stv937}, \href
  {http://adsabs.harvard.edu/abs/2015MNRAS.451...34R} {451, 34}

\bibitem[\protect\citeauthoryear{{Ro{\v s}kar}, {Teyssier}, {Agertz},
  {Wetzstein}  \& {Moore}}{{Ro{\v s}kar} et~al.}{2014}]{Roskar:2014aa}
{Ro{\v s}kar} R.,  {Teyssier} R.,  {Agertz} O.,  {Wetzstein} M.,   {Moore} B.,
  2014, \mn@doi [\mnras] {10.1093/mnras/stu1548}, \href
  {http://adsabs.harvard.edu/abs/2014MNRAS.444.2837R} {444, 2837}

\bibitem[\protect\citeauthoryear{{Rupke}, {Veilleux}  \& {Sanders}}{{Rupke}
  et~al.}{2005}]{Rupke:2005aa}
{Rupke} D.~S.,  {Veilleux} S.,   {Sanders} D.~B.,  2005, \mn@doi [\apjs]
  {10.1086/432889}, \href {http://adsabs.harvard.edu/abs/2005ApJS..160..115R}
  {160, 115}

\bibitem[\protect\citeauthoryear{{Schroeder}, {Mesinger}  \&
  {Haiman}}{{Schroeder} et~al.}{2013}]{Schroeder:2013aa}
{Schroeder} J.,  {Mesinger} A.,   {Haiman} Z.,  2013, \mn@doi [\mnras]
  {10.1093/mnras/sts253}, \href
  {http://adsabs.harvard.edu/abs/2013MNRAS.428.3058S} {428, 3058}

\bibitem[\protect\citeauthoryear{{Shapiro}, {Giroux}  \& {Babul}}{{Shapiro}
  et~al.}{1994}]{Shapiro:1994aa}
{Shapiro} P.~R.,  {Giroux} M.~L.,   {Babul} A.,  1994, \mn@doi [\apj]
  {10.1086/174120}, \href {http://adsabs.harvard.edu/abs/1994ApJ...427...25S}
  {427, 25}

\bibitem[\protect\citeauthoryear{{Shapiro}, {Iliev}  \& {Raga}}{{Shapiro}
  et~al.}{2004}]{Shapiro:2004aa}
{Shapiro} P.~R.,  {Iliev} I.~T.,   {Raga} A.~C.,  2004, \mn@doi [\mnras]
  {10.1111/j.1365-2966.2004.07364.x}, \href
  {http://adsabs.harvard.edu/abs/2004MNRAS.348..753S} {348, 753}

\bibitem[\protect\citeauthoryear{{Shapiro}, {Iliev}, {Alvarez}  \&
  {Scannapieco}}{{Shapiro} et~al.}{2006}]{Shapiro:2006aa}
{Shapiro} P.~R.,  {Iliev} I.~T.,  {Alvarez} M.~A.,   {Scannapieco} E.,  2006,
  \mn@doi [\apj] {10.1086/506242}, \href
  {http://adsabs.harvard.edu/abs/2006ApJ...648..922S} {648, 922}

\bibitem[\protect\citeauthoryear{{Shimabukuro} \& {Semelin}}{{Shimabukuro} \&
  {Semelin}}{2017}]{Shimabukuro:2017aa}
{Shimabukuro} H.,  {Semelin} B.,  2017, \mn@doi [\mnras]
  {10.1093/mnras/stx734}, \href
  {http://adsabs.harvard.edu/abs/2017MNRAS.468.3869S} {468, 3869}

\bibitem[\protect\citeauthoryear{{Somerville}}{{Somerville}}{2002}]{Somerville:2002aa}
{Somerville} R.~S.,  2002, \mn@doi [\apjl] {10.1086/341444}, \href
  {http://adsabs.harvard.edu/abs/2002ApJ...572L..23S} {572, L23}

\bibitem[\protect\citeauthoryear{{Springel}}{{Springel}}{2005}]{Springel:2005aa}
{Springel} V.,  2005, \mn@doi [\mnras] {10.1111/j.1365-2966.2005.09655.x},
  \href {http://adsabs.harvard.edu/abs/2005MNRAS.364.1105S} {364, 1105}

\bibitem[\protect\citeauthoryear{{Springel} \& {Hernquist}}{{Springel} \&
  {Hernquist}}{2003}]{Springel:2003aa}
{Springel} V.,  {Hernquist} L.,  2003, \mn@doi [\mnras]
  {10.1046/j.1365-8711.2003.06206.x}, \href
  {http://adsabs.harvard.edu/abs/2003MNRAS.339..289S} {339, 289}

\bibitem[\protect\citeauthoryear{{Srisawat}}{{Srisawat}}{2016}]{Srisawat:2016aa}
{Srisawat} C.,  2016, PhD thesis, University of Sussex

\bibitem[\protect\citeauthoryear{{Teyssier}}{{Teyssier}}{2002}]{Teyssier:2002aa}
{Teyssier} R.,  2002, \mn@doi [\aap] {10.1051/0004-6361:20011817}, \href
  {http://adsabs.harvard.edu/abs/2002A%26A...385..337T} {385, 337}

\bibitem[\protect\citeauthoryear{{Toro}, {Spruce}  \& {Speares}}{{Toro}
  et~al.}{1994}]{Toro:1994aa}
{Toro} E.~F.,  {Spruce} M.,   {Speares} W.,  1994, \mn@doi [Shock Waves]
  {10.1007/BF01414629}, \href
  {http://adsabs.harvard.edu/abs/1994ShWav...4...25T} {4, 25}

\bibitem[\protect\citeauthoryear{{Truelove}, {Klein}, {McKee}, {Holliman},
  {Howell}  \& {Greenough}}{{Truelove} et~al.}{1997}]{Truelove:1997aa}
{Truelove} J.~K.,  {Klein} R.~I.,  {McKee} C.~F.,  {Holliman} II J.~H.,
  {Howell} L.~H.,   {Greenough} J.~A.,  1997, \mn@doi [\apjl] {10.1086/310975},
  \href {http://adsabs.harvard.edu/abs/1997ApJ...489L.179T} {489, L179}

\bibitem[\protect\citeauthoryear{{Weinberg}, {Hernquist}  \& {Katz}}{{Weinberg}
  et~al.}{1997}]{Weinberg:1997aa}
{Weinberg} D.~H.,  {Hernquist} L.,   {Katz} N.,  1997, \mn@doi [\apj]
  {10.1086/303683}, \href {http://adsabs.harvard.edu/abs/1997ApJ...477....8W}
  {477, 8}

\bibitem[\protect\citeauthoryear{{Weiner} et~al.,}{{Weiner}
  et~al.}{2009}]{Weiner:2009aa}
{Weiner} B.~J.,  et~al., 2009, \mn@doi [\apj] {10.1088/0004-637X/692/1/187},
  \href {http://adsabs.harvard.edu/abs/2009ApJ...692..187W} {692, 187}

\bibitem[\protect\citeauthoryear{{Weinmann}, {Macci{\`o}}, {Iliev}, {Mellema}
  \& {Moore}}{{Weinmann} et~al.}{2007}]{Weinmann:2007aa}
{Weinmann} S.~M.,  {Macci{\`o}} A.~V.,  {Iliev} I.~T.,  {Mellema} G.,   {Moore}
  B.,  2007, \mn@doi [\mnras] {10.1111/j.1365-2966.2007.12279.x}, \href
  {http://adsabs.harvard.edu/abs/2007MNRAS.381..367W} {381, 367}

\bibitem[\protect\citeauthoryear{{Wyithe} \& {Bolton}}{{Wyithe} \&
  {Bolton}}{2011}]{Wyithe:2011aa}
{Wyithe} J.~S.~B.,  {Bolton} J.~S.,  2011, \mn@doi [\mnras]
  {10.1111/j.1365-2966.2010.18030.x}, \href
  {http://adsabs.harvard.edu/abs/2011MNRAS.412.1926W} {412, 1926}

\bibitem[\protect\citeauthoryear{{Wyithe} \& {Loeb}}{{Wyithe} \&
  {Loeb}}{2006}]{Wyithe:2006aa}
{Wyithe} J.~S.~B.,  {Loeb} A.,  2006, \mn@doi [\nat] {10.1038/nature04748},
  \href {http://adsabs.harvard.edu/abs/2006Natur.441..322W} {441, 322}

\bibitem[\protect\citeauthoryear{{Zahn}, {Lidz}, {McQuinn}, {Dutta},
  {Hernquist}, {Zaldarriaga}  \& {Furlanetto}}{{Zahn}
  et~al.}{2007}]{Zahn:2007aa}
{Zahn} O.,  {Lidz} A.,  {McQuinn} M.,  {Dutta} S.,  {Hernquist} L.,
  {Zaldarriaga} M.,   {Furlanetto} S.~R.,  2007, \mn@doi [\apj]
  {10.1086/509597}, \href {http://adsabs.harvard.edu/abs/2007ApJ...654...12Z}
  {654, 12}

\makeatother
\end{thebibliography}


\appendix

%

\section{Cold Accretion Streams}
\label{sec:cold_acc_streams}

\begin{figure}
	\includegraphics[width=\columnwidth]{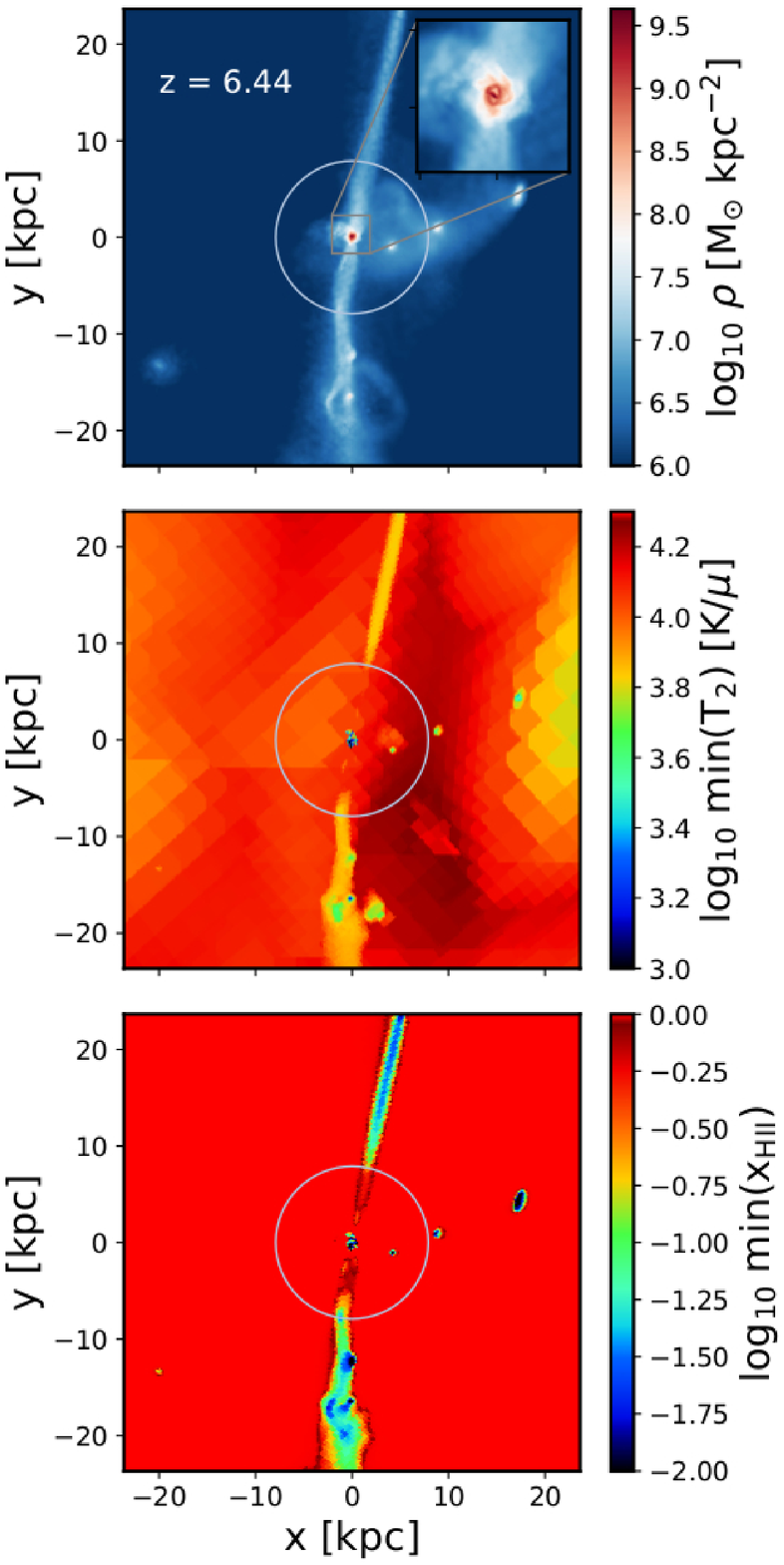}
\caption{ Projections of a 40 kpc$^{3}$ volume surrounding one of our RT2 haloes at $z \sim 6.4$ with mass $\mathrm{M}_{\mathrm{vir}} \sim 3 \times 10^{9}\ M_{\odot} h^{-1}$ and virial radius $R_{\mathrm{vir}} \sim 8$ kpc. \emph{Top panel:} Gas column density, showing the central galaxy and accretion streams. We zoom-in on the galaxy and re-project it at our full resolution. The steel-blue circle denotes the boundary of the virial sphere. \emph{Middle panel:} Minimum temperature in units of $\mathrm{K}/ \mu$, calculated by ray-tracing the volume with 512$^{2}$ rays. As gas passes the virial radius, it is shock heated to the virial temperature of the halo. \emph{Bottom panel:} Same as above, but for hydrogen ionization fraction, showing dense neutral cores within the accretion streams. This gas is sufficiently dense to self-shield from the background UV radiation. }
\label{fig:rt2_halo_self_shielding}
\end{figure}

Cold (neutral) accretion streams can form when gas reaches sufficiently high densities that it is able to self-shield from ionizing UV radiation (where recombinations balance out ionizations and equilibrium is achieved). Resolving these phenomena in fully coupled radiation hydrodynamical simulations is expensive, as one must self-consistently account for the hydrodynamical response to feedback whilst capturing these small scales in both galaxies and filaments. While some studies focus on the accurate treatment of the RT at the expense of capturing feedback \citep{Nakamoto:2001aa, Razoumov:2006aa, Ciardi:2003ab, McQuinn:2007aa, Finlator:2009aa, Aubert:2010aa}, others ignore the dynamics of the gas altogether \citep{Iliev:2006aa}, using a semi-analytical approach to calibrate the sub-grid response \citep{Iliev:2007ab}.

More recently, improvements in algorithms and the steady increase in available computing power has lead to the first fully self-consistent radiation-hydrodynamical simulations of galaxy formation to resolve this process \citep{Rosdahl:2012aa, Pawlik:2015aa, Ocvirk:2016aa}. To demonstrate the existence of self-shielding in our models, we focus on a single halo with mass $\mathrm{M}_{\mathrm{vir}} \sim 3 \times 10^{9}\ M_{\odot} h^{-1}$ and virial radius $R_{\mathrm{vir}} \sim 8$ kpc at $z \sim 6.4$ in our fiducial simulation. We show projections of gas column density (FFT convolved map), minimum temperature, and minimum hydrogen ionization fraction along rays cast through a sub-volume of 40 kpc$^{3}$ surrounding the halo (Fig.~\ref{fig:rt2_halo_self_shielding}, top to bottom respectively). The steel-blue circle annotates the boundary of the virial sphere; while in the top panel inset, we show a zoom-in of the central galaxy, re-sampled at the full resolution. In the centre and bottom panel, we see the accretion streams have a (relative) cold neutral core, which becomes disrupted as it passes the virial sphere (shock heated to the virial temperature). These streams allow the transport of cold, neutral gas from the IGM to the halo, where it may cool and form stars (provided it continues to self-shield once reaching the halo centre).

\begin{figure}
	\includegraphics[width=\columnwidth]{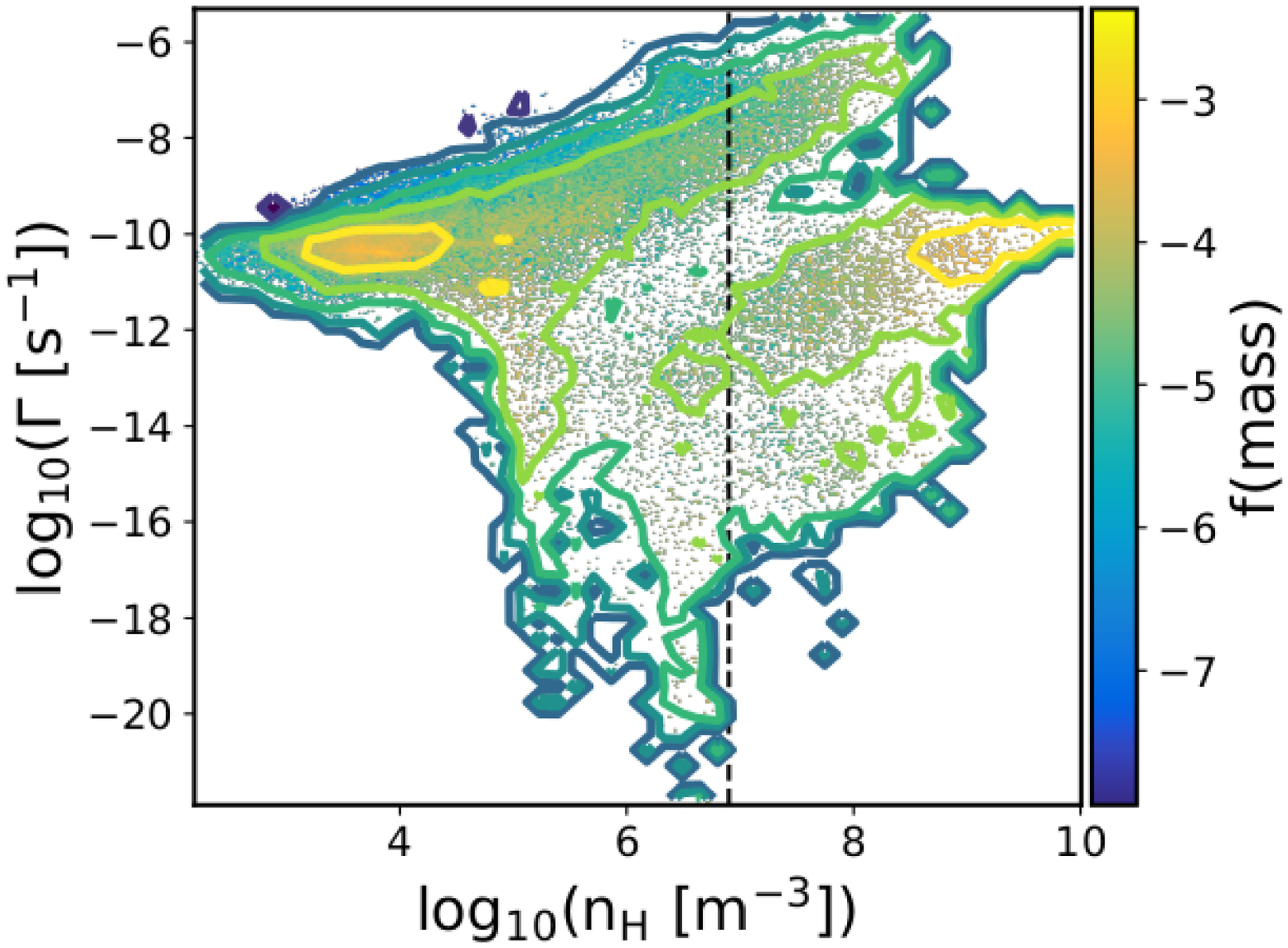}
\caption{ Mass-weighted photoionization rate-number density 2D histogram of the same volume shown in Fig.~\ref{fig:rt2_halo_self_shielding}. The black dashed line denotes the density criterion for star formation, $n_{*}$. As the density increases, the photoionization rate drops substantially, as gas is able to self-shield from ionizing UV and balance recombinations and ionizations. The photoionization rate increases at densities $n_{\mathrm{H}} > n_{*}$, as stars forming in the high density peaks, preferentially ionizing these regions. }
\label{fig:rt2_halo_self_shielding_Gamma_nH}
\end{figure}

The effect of self-shielding is made clearer by visualizing the distribution of cells in photoionization-density space, as shown in Fig.~\ref{fig:rt2_halo_self_shielding_Gamma_nH}. The black dashed line denotes the density criterion for star formation, $n_{*}$, while the colour-bar represents the fraction by mass of cells that occupy this state. As the gas density increases, the photoionization rate decreases substantially due to self-shielding. It then increases at densities $n_{\mathrm{H}} > n_{*}$ as these regions under photoionization from the inside-out in concordance with our reionization geometry (see Figs.~\ref{fig:xHII_mw_proj} and~\ref{fig:reion_hist}). This effect is also seen in \cite{Ocvirk:2016aa}, however to a lesser degree due to their limited numerical resolution (and hence maximum density achieved).

\section{Low Tidal Force Model}
\label{sec:low_tidal_force}

In this appendix, we show the results of our low-tidal force model (as described in the text). As shown in Fig.~\ref{fig:fb_amr_ANN}, strong tidal interactions act to increase the scatter in $f_{\mathrm{b}}$, especially at the low-mass end. To test the predictive power of our ANN, we re-sample the tidal forces of all haloes using the PDFs shown in Fig.~\ref{fig:pdf_ftidal}, whereby we place a cut-off at the position of the mean for the largest peak. For our simulated haloes, approximately half of the total populations exhibit tidal forces below this threshold, however our aim here is to develop a simple toy model to test if our ANN provides sensible predictions (i.e. has it properly understood the relationship between tidal force and baryon fraction).

\begin{figure*}
	\includegraphics[width=\textwidth]{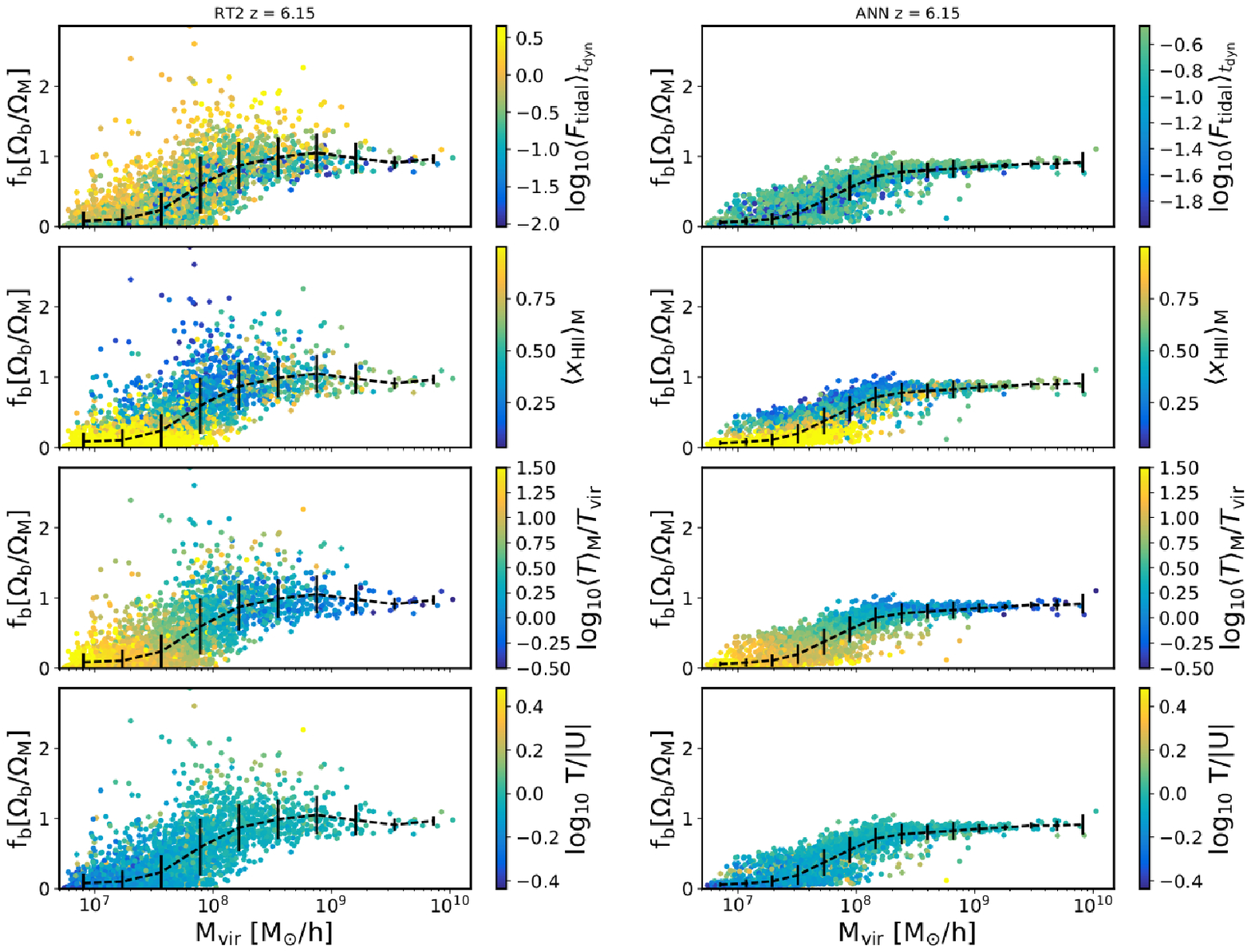}
\caption{ Same as Fig.~\ref{fig:fb_amr_ANN}, but for our low-tidal force model. We re-sample the tidal force of each halo using the PDFs shown in Fig.~\ref{fig:pdf_ftidal} and assign a new value below the position of the largest peak. Our ANN is then used to predict the halo baryon fraction for such a population, as shown in the right-hand column (note: the left-hand column is unchanged from Fig.~\ref{fig:fb_amr_ANN}, for reference). A reduction in the tidal force exerted on all haloes results in a tightening up of the scatter in $f_{\mathrm{b}}$ and an increase in our non-linear least squares fitting of $M_{\mathrm{c}}$ (see Fig.~\ref{fig:Mc_z_xHII_tidal_cutoff}, bottom panels). }
\label{fig:fb_amr_tidal_cutoff}
\end{figure*}

In Fig.~\ref{fig:fb_amr_tidal_cutoff}, we show the original halo catalogue results for our fiducial simulation (left column, as shown previously) and our ANN predictions for our test model (right column). Note, we re-scale the colour-bar for the top row for visualization purposes only. Our test model behaves as expected, leading to a significant reduction in the $\mathrm{M}_{\mathrm{vir}}$-$f_{\mathrm{b}}$ relation and also a rise in the characteristic mass (Fig.~\ref{fig:Mc_z_xHII_tidal_cutoff}, bottom panels). As discussed in the text, \cite{Okamoto:2008aa} use a similar approach to reduce scatter in their constraints, whereby they remove all haloes that have undergone recent merger activity. Exclusion of such objects leads to an increases in the numerical values for $M_{\mathrm{c}}$, as we have shown.

\bsp	
\label{lastpage}
\end{document}